\documentclass[aps,pre,superscriptaddress,twocolumn]{revtex4-2}

\usepackage{graphicx,chemarr,color,hyperref}
\usepackage[version=4]{mhchem}

\usepackage{amssymb,amsmath,textgreek}

\begin{document}

\title{Multicellular sensing at a feedback-induced critical point}

\author{Michael Vennettilli}
\affiliation{Department of Physics and Astronomy, Purdue University, West Lafayette, Indiana 47907, USA}
\affiliation{Department of Physics and Astronomy, University of Pittsburgh, Pittsburgh, Pennsylvania 15260, USA}

\author{Amir Erez}
\affiliation{Department of Molecular Biology, Princeton University, Princeton, NJ 08544, USA}
\affiliation{The Racah Institute of Physics, Hebrew University, Jerusalem 91904, Israel}

\author{Andrew Mugler}
\email{amugler@purdue.edu}
\affiliation{Department of Physics and Astronomy, Purdue University, West Lafayette, Indiana 47907, USA}
\affiliation{Department of Physics and Astronomy, University of Pittsburgh, Pittsburgh, Pennsylvania 15260, USA}

\begin{abstract}
Feedback in sensory biochemical networks can give rise to bifurcations in cells' behavioral response. These bifurcations share many properties with thermodynamic critical points. Evidence suggests that biological systems may operate near these critical points, but the functional benefit of doing so remains poorly understood. Here we investigate a simple biochemical model with nonlinear feedback and multicellular communication to determine if criticality provides a functional benefit in terms of the ability to gain information about a stochastic chemical signal. We find that when signal fluctuations are slow, the mutual information between the signal and the intracellular readout is maximized at criticality, because the benefit of high signal susceptibility outweighs the detriment of high readout noise. When cells communicate, criticality gives rise to long-range correlations in readout molecule number among cells. Consequently, we find that communication increases the mutual information between a given cell's readout and the spatial average of the signal across the population. Finally, we find that both with and without communication, the sensory benefits of criticality compete with critical slowing down, such that the information rate, as opposed to the information itself, is minimized at the critical point. Our results reveal the costs and benefits of feedback-induced criticality for multicellular sensing.
\end{abstract}

\maketitle

\section{Introduction}

Cells need to reliably sense their environment to survive and coordinate behavior. Many studies have investigated the precision of sensory tasks, such as detecting concentrations of a single molecular species \cite{berg1977chemoreception, kaizu2014revisited, ngampruetikorn2020energy}, concentrations of multiple molecular species \cite{carballopacheco2019xtalk, mora2015spurious}, and concentration gradients \cite{endres2008gradient, hu2010gradients}.
Much of those works focused on linear networks in single cells. Yet, it is well known that biological systems use nonlinear feedback to process signals, and there is evidence that communication among multiple cells improves sensory precision \cite{gregor2007positional, ellison2016communication, fancher2017collective}. For example, in the developing embryo of the {\it Drosophila melanogaster} fruit fly, cell nuclei sense their position along a Bicoid (Bcd) protein gradient using Hunchback (Hb) as a genetic readout. Hb has binding sites on its own promoter region leading to positive feedback \cite{treisman1989products}, and it is also thought to be diffusively communicated from nucleus to nucleus in the embryo \cite{gregor2007positional, erdmann2009averaging}. In \textit{Vibrio fischeri} bacteria, cells sense the concentration of an autoinducer (AI). The uptake of AI is thought to cooperatively produce more AI via the lux operon, creating a nonlinear positive feedback loop, and AI is communicated diffusively among cells for the purpose of quorum sensing \cite{williams2008quorum}. Here we probe theoretically this interplay between cellular sensing, feedback, and communication.

Positive feedback generically leads to bifurcations in dynamical behavior \cite{strogatz2018nonlinear}. This raises the question of what the implications of being near a bifurcation are for biological sensing. In the presence of noise, which is ubiquitous in biochemical networks, these systems behave like thermodynamic critical systems, exhibiting characteristic features like power law scalings and critical slowing down \cite{erez2019universality, byrd2019slow}. Indeed, there is experimental evidence suggesting that some biological systems operate near criticality \cite{mora2011criticality, krotov2013morphogenesis, munoz2018criticality}. On the one hand, critical systems have divergent susceptibilities and correlation lengths, which may be beneficial for sensing. On the other hand, critical systems have large fluctuations and slow dynamics, which may be detrimental for sensing. These expected tradeoffs suggest that the costs and benefits of sensing at criticality need to be explored in a systematic way.

Here we probe the implications of criticality for the sensing of a noisy, spatially uniform chemical concentration by a population of communicating cells. We focus on a variant of Schl{\"o}gl's second model \cite{schlogl1972chemical} that incorporates linear sensing, nonlinear feedback, and communication between neighboring cells. We consider cells on a one-dimensional lattice, but we show that the model is in the mean-field Ising static universality class, and therefore we expect our steady-state results to qualitatively hold for more general geometries. In the case of a single cell, we find that critical feedback maximizes the mutual information when the ligand dynamics are slow due to the high susceptibility of the response to the input. Similarly, with multiple cells, we find that critical feedback couples with cell-cell communication to produce long-range correlations, which maximizes information about the average ligand concentration across the population. However, we find that in both cases, critical feedback results in critical slowing down, such that the information rate is minimized. We discuss the implications of these tradeoffs for several well-studied biological systems.

\section{Results}

\subsection{The Model and Its Universality Class}
\begin{figure}
	\includegraphics[width=1\columnwidth]{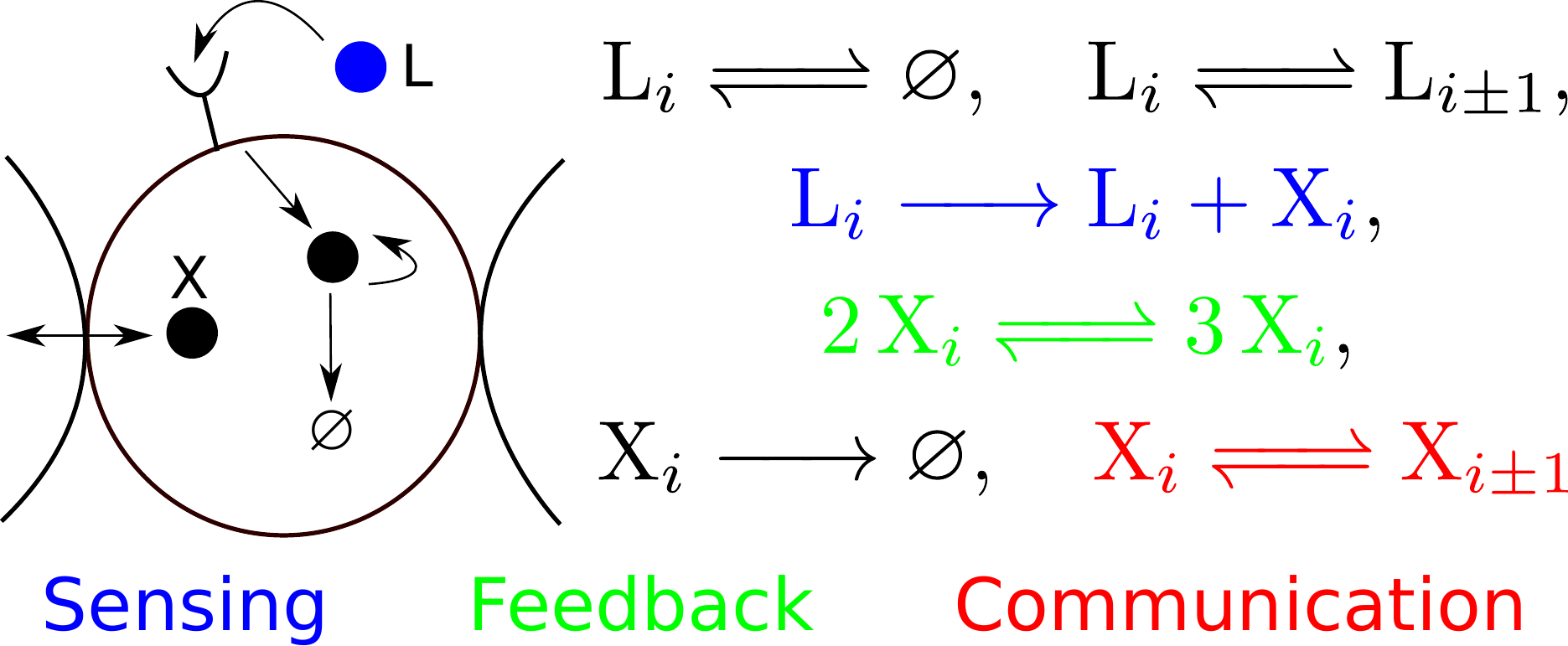}
	\caption{Illustration of the model. Each site on a 1D lattice is a well-mixed cell and its immediate environment. A ligand binds to a receptor and produces a readout molecule that is subject to nonlinear feedback and may be exchanged between cells. The ligand is spatially uniform and has its own fluctuating dynamics.}
	\label{fig:model}
\end{figure}

We consider a one-dimensional chain of sites with periodic boundary conditions, where each site corresponds to a cell and its immediate environment (Fig.\ \ref{fig:model}). We assume the chemical components within each site are well mixed. Our model generalizes Schl{\"o}gl's second model for nonlinear biochemical feedback \cite{schlogl1972chemical} to multiple coupled cells and introduces an extracellular diffusing ligand to be sensed. The number of ligand molecules at site $i$ is denoted by $\ell_i$, while the number of readout molecules in cell $i$ is denoted $x_i$. Specifically, we have the reactions
\begin{equation} \label{eq:rxns}
\begin{gathered}
\ce{L_i <=>[k^-_{\ell}][k^+_{\ell}] \varnothing},\quad \ce{L_i <=>[\gamma^{\prime}][\gamma^{\prime}] L_{i+ 1}} , \quad	\ce{L_i ->[k^+_1] X_i + L_i}, \\
\ce{X_i ->[k^-_1] \varnothing},\quad \ce{2 X_i <=>[k^+_2][k^-_2] 3 X_i}, \quad \ce{X_i <=>[\gamma][\gamma] X_{i+ 1}}.
\end{gathered}	
\end{equation}
The first two reversible reactions describe the ligand diffusively entering or leaving the vicinity of the $i$th cell or diffusing to that of the neighboring cell. The third reaction describes the production of a readout molecule in response to the ligand. The fourth reaction describes linear degradation of the readout. The fifth reaction subjects the readout to positive and negative nonlinear feedback. The sixth reaction describes exchanging the readout between neighboring cells. The fourth and fifth reactions are often written with additional, mediating bath species, but these are usually assumed to have a fixed concentration, and therefore here we have absorbed them into the rate constants.

The nonlinear feedback in Eq.\ \ref{eq:rxns} has its own mechanistic interpretation \cite{schlogl1972chemical}, but it also emerges as the Taylor series for a wide range of nonlinear reactions near a pitchfork bifurcation \cite{erez2019universality, byrd2019slow}. For example, this is the case if the feedback reactions were replaced with production via a Hill function in $x$ with Hill coefficient $H >1$. Eq.\ \ref{eq:rxns} is also convenient mathematically, as it can be mapped onto the mean-field Ising model with a translation and multiplicative scaling, as we now show.

In previous work, we mapped the single cell onto the mean-field Ising model by solving for the steady state probability distribution of readout molecule number \cite{erez2019universality}. This approach breaks down with multiple coupled cells because the steady state distribution is not known analytically. Instead, we extract the mapping from the deterministic dynamics \cite{bose2019criticality}. The rate equations for the readout take the form
\begin{equation}
	\frac{d x_i}{dt} = k_1^+\ell_i - k_1^+ x_i + k_2^+ x_i^2 - k_2^- x_i^3 +\gamma \nabla^2 x_i,
\end{equation}
where $\nabla^2$ is the discrete Laplace operator. The key is to perform a change of variables so that this equation of motion is in the normal form of a supercritical pitchfork bifurcation, as the relaxation dynamics to the minima of the Landau free energy of the Ising model are in this form \cite{strogatz2018nonlinear, goldenfeld1992renorm}. This is accomplished by finding where the second derivative of the right-hand side with respect to $x_i$ vanishes, and this occurs at $x_c=k_2^+/3k_2^-$. Moreover, $x_c$ controls the typical molecule number in the system, with scaling properties similar to finite-size scaling in Ising lattices. Making the substitution $x_i = x_c(m_i+1)$, where $m_i$ is an order parameter analogous to the magnetization of the Ising model, and rescaling time so that the cubic term has a coefficient of $-1/3$ gives
\begin{align}
\label{eq:m}
	\begin{split}
	&\frac{dm_i}{d\tau} = -\frac{1}{3} m_i^3 -\left[ 3 \frac{k_1^- k_2^-}{(k_2^+)^2} -1 \right]m_i \\
	 &+ \left[\frac{2}{3} -3 \frac{k_1^- k_2^-}{(k_2^+)^2} + 9 \frac{k_1^+ \ell_i (k_2^-)^2}{(k_2^+)^3}\right] + 3\frac{k_2^- \gamma}{(k_2^+)^2} \nabla^2 m_i.
	\end{split}
\end{align}
Eq.\ \ref{eq:m} describes relaxation dynamics to the minima of the Landau free energy for the Ising model, provided that we identify
\begin{gather}
\label{eq:theta}
\theta = 3 \frac{k_1^- k_2^-}{(k_2^+)^2} -1, \\
\label{eq:h}
h_i = \frac{2}{3} -3 \frac{k_1^- k_2^-}{(k_2^+)^2} + 9 \frac{k_1^+ \ell_i (k_2^-)^2}{(k_2^+)^3}
\end{gather}
as the reduced temperature and dimensionless field respectively. The bifurcation occurs when both of these parameters are set to zero \cite{strogatz2018nonlinear}. The field $h_i$ biases the distribution to either high or low molecule numbers. This can be understood by the quantitative analogy between the magnetization $m_i$ and the molecule number $x_i$. For a positive magnetic field, the magnetization will be biased towards positive values, so the molecule number will be biased towards values larger than $x_c$, and a similar analogy holds for negative fields. Because we are interested in properties of the system near the critical point, we set $h_i = 0$ (using the mean value of the ligand $\overline{\ell}=k_{\ell}^+/k_{\ell}^-$ for $\ell_i$). When the reduced temperature $\theta$ is decreased from positive to negative, the feedback becomes increasingly strong, and the distribution goes from unimodal to bimodal. Eq.\ \ref{eq:m} differs from the standard form of $\varphi^4$ theory \cite{goldenfeld1992renorm}, where the coefficient of the Laplacian term would be $1$.
Instead this coefficient depends on the exchange rate $\gamma$ of molecules from cell to cell, which is an independently tunable parameter in the biochemical model. These parameters come from the deterministic dynamics, so they only capture the large $x_c$ limit of the stochastic dynamics. We ignore corrections from small molecule numbers \cite{erez2019twocell} and purely stochastic effects \cite{zhang2010vanthoff}.

The fact that this system is in the static universality class of the mean field Ising model may appear at odds with previous work done on Schl{\"o}gl's second model extended to a spatial context \cite{brachet1981schlogl2, grassberger1981montecarlo, grassberger1982phasetransition}. The reason is that in these previous works, there is a finite occupancy per site and each molecule can only react with neighboring molecules. In our case, each molecule can interact with an arbitrary number of molecules within a single well-mixed site. This makes the mean field nature of the model apparent, and one should think of our model as a system of linearly coupled mean field systems.

\subsection{Defining the Sensory Measure}

A typical measure of sensory precision is the signal-to-noise (SNR) ratio in the cell's estimate of the ligand concentration. In Fig. \ref{fig:ssmi}A we show the SNR $\bar{x}^2/\sigma_x^2$ for a single cell as a function of $\theta$, computed using Gillespie simulations \cite{gillespie1977exact} of Eq.\ \ref{eq:rxns}. The parameter $\theta$ inversely sets the feedback strength, with $\theta \to\infty $ corresponding to no feedback and $\theta = 0$ corresponding to critical feedback. We avoid the $\theta<0$ case (feedback-induced bifurcation) since the mean and variance of a bimodal distribution are ill-suited measures in this context. We see that the SNR increases with $\theta$, meaning that criticality is worse for the SNR than having no feedback. The reason is that, for $\theta\geq 0$, the mean $\bar{x}$ is independent of $\theta$ while the variance $\sigma_x^2$ increases as $\theta$ decreases and the critical point is approached. Evidently criticality is not beneficial for sensing a mean concentration with low error.

In contrast, we hypothesize that criticality might be beneficial for sensing fluctuations around a mean concentration. This is because critical systems have large susceptibility to a biasing field. Indeed, the fluctuating ligand number $\ell_i$ appears in the effective field $h_i$ (Eq.\ \ref{eq:h}). Therefore, we hypothesize that the critical system may have maximal correlations between the ligand and readout, and therefore serve as an optimal fluctuation detector. To investigate this hypothsis, we consider the mutual information and information rate \cite{shannon1948communication} between the ligand and readout molecule numbers as our sensory measures from here on. The mutual information and information rate inherently capture correlations between two fluctuating variables (see Appendix A for a summary of results for Gaussian variables, which we will draw upon below).

\subsection{Single Cell}

We start by analyzing the Langevin equations obtained from the Kramers-Moyal expansion for our system in the limit of a single cell. Even for the single cell, the nonlinearity is an obstruction to exact analytic solutions, and therefore we linearize the equation about a deterministic steady state. When $\theta$ is negative, the distribution is bimodal; this feature cannot be captured by the linearized equation, so this approximation is only valid for $\theta>0$. We will use this approach to gain insight and then use simulations for more accurate results. When $\theta>0$ and $h=0$, there is only one deterministic fixed point; it occurs at $\overline{x} =x_c$ and is stable. For this reason, we will treat $x_c$ as a tuning parameter that sets the system size. 

\subsubsection{Mutual Information}

Letting $b(x,\ell) = k_1^+\ell + k_2^+ x^2$ and $d(x) = k_1^- x + k_2^- x^3$ denote the total birth and death terms for $x$ (Eq.\ \ref{eq:rxns}), we obtain the linearized system
\begin{equation}\label{eq:scSDEs}
	\begin{gathered}
	\dot{\delta \ell} = -k_{\ell}^- \delta \ell + \sqrt{2k_{\ell}^+} \epsilon, \\
	\dot{\delta x} = -c \delta x + k_1^+ \delta \ell + \sqrt{2 d(x_c)}\eta,
	\end{gathered}
\end{equation}
where $\epsilon$ and $\eta$ are independent, delta-correlated Gaussian white noise processes that describe the noise in production and degradation (Appendix B). The $c = \partial_x [d(x)-b(x,\ell)]_{x_c,\overline{\ell}}$ term arises from linearizing the birth and death propensities with respect to $x$. This system is a two-dimensional Ornstein-Uhlenbeck process and may be solved analytically using matrix exponentials and It{\^o}'s lemma. This is done in Appendix B and we find that the steady-state mutual information between $x$ and $\ell$ is
\begin{equation}
\label{eq:Iss}
	I=\frac{1}{2} \log\left(1 + \frac{(k_1^+)^2 \overline{\ell} c}{d(x_c)(k_{\ell}^- + c)^2 +(k_1^+)^2 k_{\ell}^- \overline{\ell}}\right).
\end{equation}
We expect the system to track ligand fluctuations best when the ligand dynamics are slow ($k_{\ell}^-/k_1^- \ll 1$). Writing Eq.\ \ref{eq:Iss} in this limit and in terms of the reduced temperature (Eq.\ \ref{eq:theta}), we find
\begin{equation}
\label{eq:Itheta}
	I = \frac{1}{2} \log \left(1 +\frac{x_c}{\overline{\ell}} \frac{(3\theta +1)^2}{3 \theta(3\theta +4)}\right).
\end{equation}
This result has a finite limit when $\theta \rightarrow \infty$ (no feedback) and diverges as $\theta \to 0$ (critical feedback). This suggests that indeed, the mutual information between ligand and readout is maximized at criticality under these assumptions. Moreover, the role of $x_c$ as the system size is apparent, with the mutual information increasing approximately as $\log (x_c/\bar{l})$.

To understand intuitively why the information is maximized at criticality, we can write Eq.\ \ref{eq:Itheta} in terms of new variables: the susceptibility $\chi = (\partial m/\partial h)_{h=0}$ and the variance $\sigma_x^2$ with $\ell$ fixed to its mean value. The susceptibility is obtained directly from Eq.\ \ref{eq:m}, which for the single cell in terms of $\theta$ and $h$ reads $0 = -m^3/3-\theta m+h$ at steady state. Differentiating with respect to $h$ and evaluating at $h=0$ yields $0 = -(m^2)_{h=0}\chi - \theta\chi+1$, or $\chi = 1/\theta$, where we have recognized that $m=0$ for $h=0$ and $\theta>0$. The variance is $\sigma_x^2= d(x_c)/c$ (Appendix B), or in terms of the reduced temperature, $\sigma_x^2 = x_c(\theta+4/3)/\theta$. In terms of these variables, Eq.\ \ref{eq:Itheta} reads
\begin{equation}
\label{eq:Ichi}
	I =\frac{1}{2} \log\left(1 + \frac{(\chi/3 +1)^2}{\overline{\ell} \sigma_x^2 /x_c^2}\right).
\end{equation}
We see from the expressions for $\chi$ and $\theta$ that both the susceptibility and noise diverge at the critical point like $1/\theta$. However, we see from Eq.\ \ref{eq:Ichi} that the information scales monotonically with the square of the susceptibility in the numerator, but only the first power of the variance in the denominator. This implies that the benefit of high susceptibility outweighs the detriment of high noise, making criticality optimal for this type of information transmission.

\begin{figure}
	\includegraphics[width=1\columnwidth]{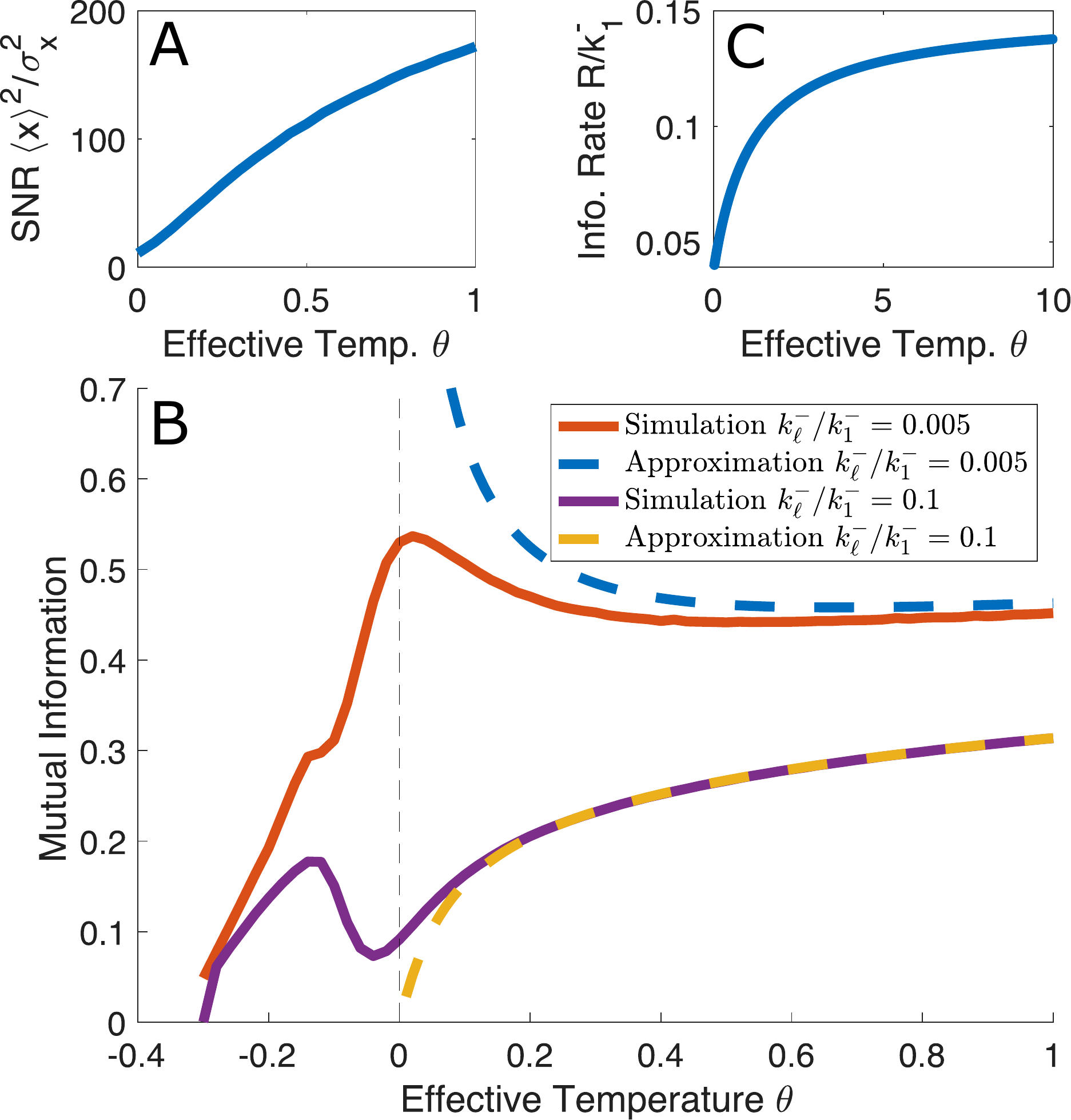}
	\caption{Sensory information for a single cell.  (A) Signal-to-noise ratio for the readout in the slow ligand regime with $k_{\ell}^-/k_1^- = 5\times 10^{-3}$. (B) Mutual information between ligand and readout. Blue (yellow): Linear noise approximation Eq.\ \ref{eq:Itheta} (Eq.\ \ref{eq:Ithetafull}) with $k_{\ell}^-/k_1^- = 0$ ($0.1$). Red (purple): Gillespie simulations with $k_{\ell}^-/k_1^- = 5\times 10^{-3}$ ($0.1$). (C) Information rate between ligand and readout using linear noise approximation with $k_{\ell}^-/k_1^- = 5\times 10^{-3}$. All curves have $h=0$, $x_c = 10^3$, and $\overline{\ell}=500$.}
	\label{fig:ssmi}
\end{figure}

The linear noise approximation breaks down at the critical point, and therefore we use Gillespie simulations \cite{gillespie1977exact} to check our results. The simulations also allow us to probe the bimodal regime ($\theta < 0$). The results are shown in Fig. \ref{fig:ssmi}B. In the slow ligand limit, $k_{\ell}^-/k_1^- \ll 1$, shown in the red curve, we see that the divergence at the critical point is rounded off to a global maximum just above the critical point due to the finite size of the system.

The location of the global maximum changes discontinuously as the timescale ratio $k_{\ell}^-/k_1^-$ is increased. As soon as $k_{\ell}^-/k_1^-$ becomes nonzero, the mutual information vanishes at the critical point in the linear noise approximation (Appendix B, Eq.\ \ref{eq:Ithetafull}), as seen for the yellow curve in Fig.\ \ref{fig:ssmi}B. In the simulations, as $k_{\ell}^-/k_1^-$ is increased, the ``blip" on the red curve for $\theta < 0$ separates and forms another peak, with a minimum appearing between the two local maxima. The height of the peak decreases until it dips below the asymptotic value at large $\theta$ and later disappears entirely, as seen for the purple curve in Fig.\ \ref{fig:ssmi}B. Thus, when the ligand timescales are sufficiently fast, the mutual information is maximized in the absence of feedback. It is possible to estimate where criticality ceases to be highly informative using the linear noise approximation (Appendix B, Eq.\ \ref{eq:Ithetafull}). Under this approximation, the system with feedback cannot outperform the system without feedback if $k_{\ell}^-/k_1^- \geq (3\sqrt{3}-5)/8 \approx 0.025$. This suggests that a timescale separation of nearly two orders of magnitude is necessary to benefit from feedback.

We can also probe the effect of changing the system sizes $\overline{\ell}$ and $x_c$. Although we could use simulations, the results depend on a choice of $k_{\ell}^-/k_1^-$, and the peak may not exist if this ratio is too large. We study the $k_{\ell}^-/k_1^-\rightarrow 0$ limit numerically by writing the joint distribution as $P(x|\ell)P(\ell)$, where $P(\ell)$ is a Poisson distribution with mean $\overline{\ell}$ and $P(x|\ell)$ is computed from the master equation assuming that $\ell$ is constant. We find that, even when the timescale separation is infinitely large, the peak at the critical point can vanish if either of the molecule numbers are sufficiently small. In this case, the mutual information increases monotonically with $\theta$.

\subsubsection{Information rate}

Cells respond to their environment and make life-or-death decisions in real time, and do not have the luxury to wait until all possible information has been collected, leading some to argue that the information rate is more relevant than the information itself \cite{tostevin2009trajectories, meijers2019infocrit}. The information rate, $R$, is the asymptotic rate of change of the mutual information between trajectories of the input and output \cite{munakata2006dynamic}. Concretely, we sample the input and output at discrete times, usually multiples of some $\delta t>0$, and construct vectors of the samples $\vec{i}_N=\{i(0), i(\delta t), ..., i((N-1)\delta t) \}$ and $\vec{o}_N=\{o(0), o(\delta t), ..., o((N-1)\delta t) \}$. We regard $\vec{i}_N$ and $\vec{o}_N$ as random variables and compute the mutual information between them. Finally, the information rate is computed as
\begin{equation}
	R(i,o) = \lim\limits_{\delta t\rightarrow 0^+} \left(\lim\limits_{N\rightarrow \infty} \frac{I(\vec{i}_N,\vec{o}_N)}{N\delta t} \right).
\end{equation}
It is possible to compute the information rate between the ligand and readout fluctuations under the linear noise approximation for our system. In Appendix C, we show that the result is
\begin{equation}
\label{eq:R}
	R = \frac{\pi k_{\ell}^-}{2} \left( \sqrt{1 + \frac{x_c}{3\overline{\ell}} \frac{k_1^-}{k_{\ell}^-} \left[ \frac{(1+3\theta)^2}{4 +7\theta + 3\theta^2}\right]} -1\right).
\end{equation}
Eq.\ \ref{eq:R} vanishes when the ligand degradation rate vanishes. This makes sense as the ligand degradation rate sets the timescale for the ligand dynamics, and the ligand is the signal that the readout is trying to track. Eq.\ \ref{eq:R} is a monotonically increasing function of the temperature for $\theta \geq 0$ (Fig.\ \ref{fig:ssmi}C). This suggests that the information rate has a global minimum at criticality and is maximized without feedback, at least for $\theta\geq 0$. This reveals an interesting tradeoff: the steady-state mutual information decreases as ligand rates increase and is maximized at criticality when the ligand rates are slow, while the information rate is maximized when the ligand rates are fast and is minimized at criticality. Discretely sampled trajectories are high-dimensional in this problem, which is an obstruction to efficient and accurate simulation via the Gillespie algorithm \cite{meijers2019infocrit}. Although the instantaneous distribution for the ligand and readout is two-dimensional, the joint distribution for sampling dynamic trajectories at $N$ points is $2N$-dimensional. However, we still expect the information rate to decrease as $\theta$ decreases towards zero, as positive feedback is known to increase response time \cite{alon2007systems} and critical slowing down will make this more extreme.

\subsection{Multiple Cells with Communication}

\subsubsection{Long-range Correlations}

\begin{figure}
	\includegraphics[width=1\columnwidth]{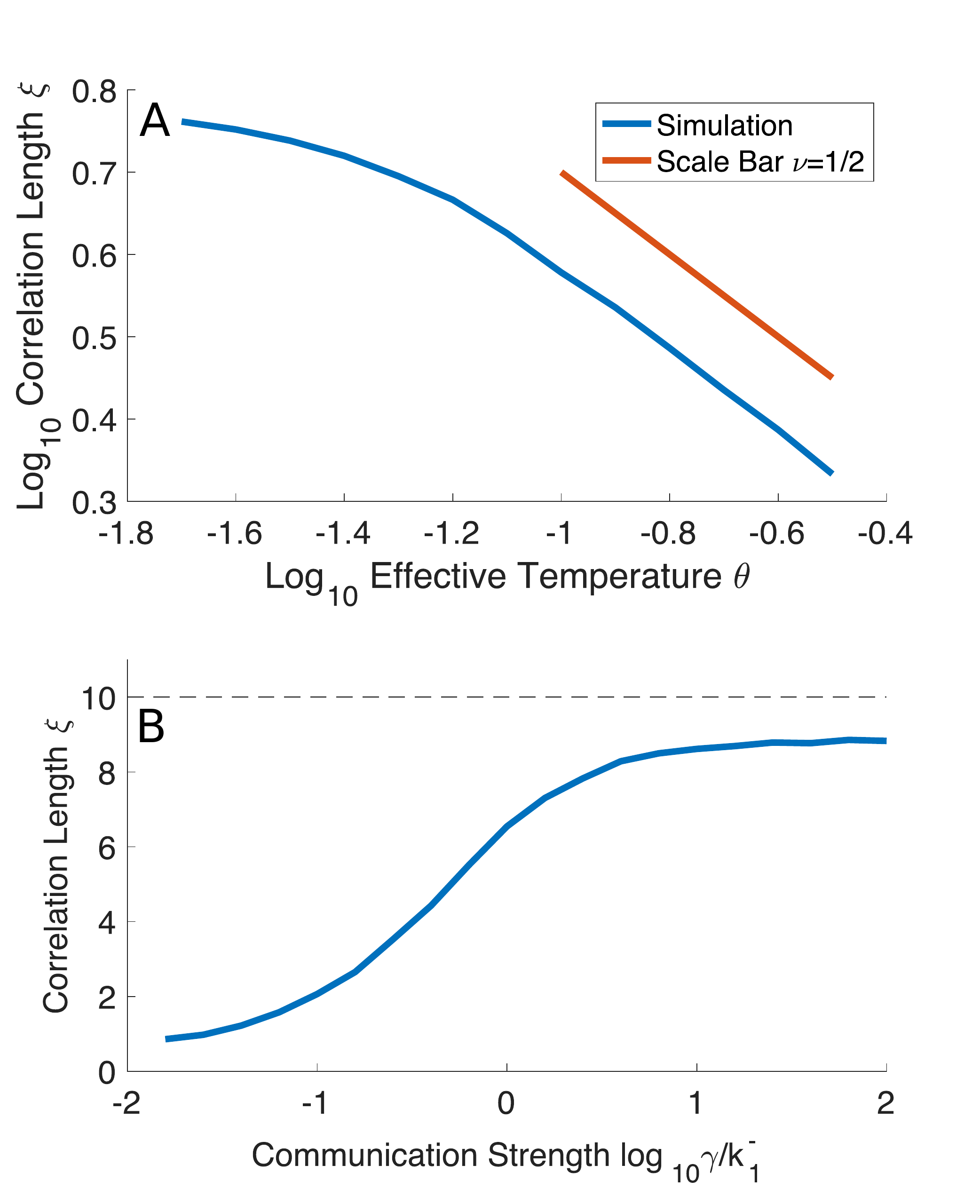}
	\caption{ Long-range correlations in the multicellular system.  (A) Correlation length $\xi$, numerically integrated from simulations using the trapezoidal rule, scales with reduced temperature $\theta$ with mean-field exponent $\nu = 1/2$. Rolloff at $\theta=0$ is due to finite size effects. Here $k_{\ell}^-/k_1^- = \gamma'/k_1^-= 5\times 10^{-3}$, $\gamma/k_1^-=1$, $h_i=0$, $\overline{\ell}=150$, $x_c=300$, and $N=20$ cells. (B) Critical correlation length increases with communication strength $\gamma$, approaching system size $N/2$. Parameters as in A with $\theta=0$.}
	\label{fig:longCorr}
\end{figure}

In previous work \cite{erez2019universality}, we established that the single cell had the same static critical exponents as the mean field Ising model. In this section, we will show that the multicellular system inherits the mean field exponent for the the correlation length, $\nu=1/2$, at least for $\theta>0$, and therefore exhibits long-range correlations among cells. To find the critical exponent, we compute the spatial correlation function between different cells' readouts using Gillespie simulations \cite{gillespie1977exact}. Here, we use the trapezoidal rule to integrate the correlations along both sides of the multicellular chain and then average the two results. The resulting correlation length $\xi$ is plotted as a function of $\theta$ in Fig.\ \ref{fig:longCorr}A. When $\theta$ is very close to the critical point, finite size $x_c$ effects become important and lead to rounding off. However, we can see the true scaling by moving away from the critical point so that the finite size effects become less limiting. Here we see that $\xi\sim \theta^{-1/2}$, or $\nu=1/2$, as in the mean field Ising model \cite{goldenfeld1992renorm}.

Next, we looked at the effect of communication strength $\gamma$ on the correlation length. We expect that the correlation length at $\theta = 0$ should roughly approach the system size $N/2$ in the strong communication limit $\gamma\rightarrow\infty$, where $N$ is the number of cells, and the factor of $2$ is due to the periodic boundary condition. The result is shown in Fig.\ \ref{fig:longCorr}B for $N=20$ cells, and we see that the correlation length indeed approaches $N/2$ as $\gamma$ becomes large.

\subsubsection{Sensory Information}

To compute the sensory information for the multicellular system, we proceed with the Langevin equations as we did with the single cell.  The key difference here is that there are discrete Laplacian terms and noise terms for diffusion left or right for each site. The linearized Langevin equations become
\begin{equation}
	\begin{aligned}
	\dot{\delta \ell}_i = &-k_{\ell}^- \delta \ell_i + \gamma' \nabla^2 \delta \ell_i + \sqrt{2 k_{\ell}^+} \epsilon_{i,C} \\
	&+\sqrt{\gamma' \overline{\ell}}(\epsilon_{i-1,R} + \epsilon_{i+1,L} -\epsilon_{i,L} -\epsilon_{i,R}), \\
	\dot{\delta x}_i = &-c\delta x_i + k_1^+ \delta \ell_i + \gamma \nabla^2 \delta x_i +\sqrt{2d(x_c)}\eta_{i,C} \\
	&+\sqrt{\gamma x_c} (\eta_{i-1,R} + \eta_{i+1,L} -\eta_{i,L} -\eta_{i,R}),
	\end{aligned}
\end{equation}
where the $\eta$'s and $\epsilon$'s are independent white noise processes (Appendix D). The terms with subscript $L$ and $R$ describe the noise in hopping to the left and right respectively, while the $C$ terms describe the changes due to the chemical processes. The linearized system is still a multi-dimensional Ornstein-Uhlenbeck process. One can obtain closed form results for the matrix exponentials and covariance matrix by using translational invariance. These calculations are done in Appendix D.

With multiple cells, there are several possible measures of sensory information to consider. Because we aim to focus on the actions that an individual cell can take, we take the output to be the readout in a single cell, $\delta x_i$. However, there are multiple possibilities for the sensory input: the local ligand fluctuations $\delta \ell_i$, the ligand fluctuations at all cell locations $\vec{\delta \ell}$, or the spatial average of the ligand fluctuations
\begin{equation}
	\delta L = \frac{1}{N} \sum\limits_{i=1}^N \delta \ell_i.
\end{equation}
We compute the mutual information between each of these inputs and the output in Appendix E. In the slow ligand limit, we find that all of them diverge at criticality as in the single-cell case, and therefore we focus on the effect of tuning the communication rate $\gamma$.

We find that the first mutual information $I(\delta x_i, \delta \ell_i)$ decreases as the communication between cells is increased (Appendix E).
This makes intuitive sense because $I(\delta x_i, \delta \ell_i)$ is the information between local ligand fluctuations and local readout. Communication among cells only mixes one cell's readout with the other cells' readouts. The other cells' readouts report on distant ligand fluctuations, which are uncorrelated with the local ligand fluctuations because the ligand molecules at different sites are independent Poisson random variables. Therefore $I(\delta x_i, \delta \ell_i)$ is maximized in the absence of cell-cell communication.

We also find that the second mutual information $I(\delta x_i,\vec{\delta \ell})$ decreases as communication between cells is increased (Appendix E). For this sensory task, the cell must be able to reliably encode the entire spatial profile $\vec{\delta \ell}$ into a single, fluctuating variable $\delta x_i$. On the one hand, we expect cell-cell communication to help in this task because communication transmits information about distant ligand fluctuations sensed by the other cells. One the other hand, as in the previous case, communication obscures the information that the cell directly obtains about its local environment $\delta \ell_i$, which is one of the components of $\vec{\delta \ell}$. Evidently the latter effect dominates.
The mutual information is maximized in the absence of communication, where it can sense its immediate environment reliably.

We find that the third mutual information $I(\delta x_i, \delta L)$ increases as the communication between cells is increased. Like the spatial profile $\vec{\delta \ell}$, the spatial average of the ligand fluctuations $\delta L$ also contains global information. However, unlike in the previous case, for this sensory measure it is not detrimental that the single cell's readout combines local and global environmental information. This is because here the cell only senses average environmental changes. An increase in its readout is correlated with an increase in the ligand somewhere, and for this task it does not matter where. It is worth pointing out that a cell is most strongly correlated with its immediate neighborhood. As $N$ increases, its neighborhood contributes less strongly to the average, and the mutual information monotonically decreases.

\begin{figure}
	\includegraphics[width=1\columnwidth]{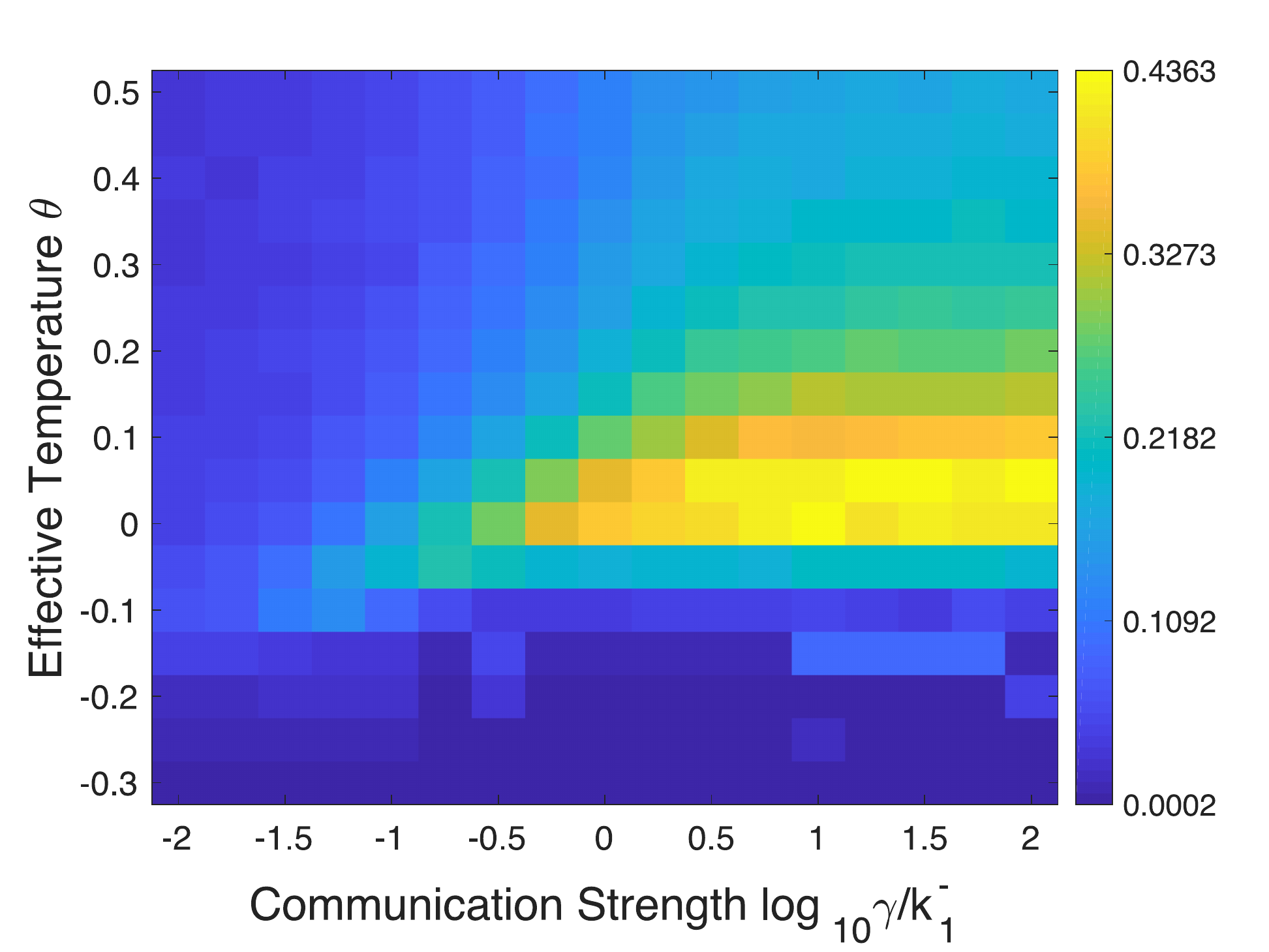}
	\caption{Mutual information between a single cell's readout and the spatial average of the ligand fluctuations. This was obtained from Gillespie simulations with $N=10$, $h_i=0$, $x_c = 300$, $\overline{\ell}=150$, and $k_{\ell}^-/k_1^- = \gamma'/k_{\ell}^- = 5\times 10^{-3}$. This is maximized near criticality as the communication strength increases.}
	\label{fig:multimi}
\end{figure}

To verify the linear noise approximation, we again use Gillespie simulations to probe the exact behavior of $I(\delta x_i, \delta L)$. The mutual information in the slow ligand regime is shown in Fig. \ref{fig:multimi}. Like in the single cell case, it is largest near the critical point when the communication strength is fixed. For $\theta \geq 0$, the mutual information increases with the communication rate, as expected from the linear noise approximation. 

\begin{figure}
	\includegraphics[width=1\columnwidth]{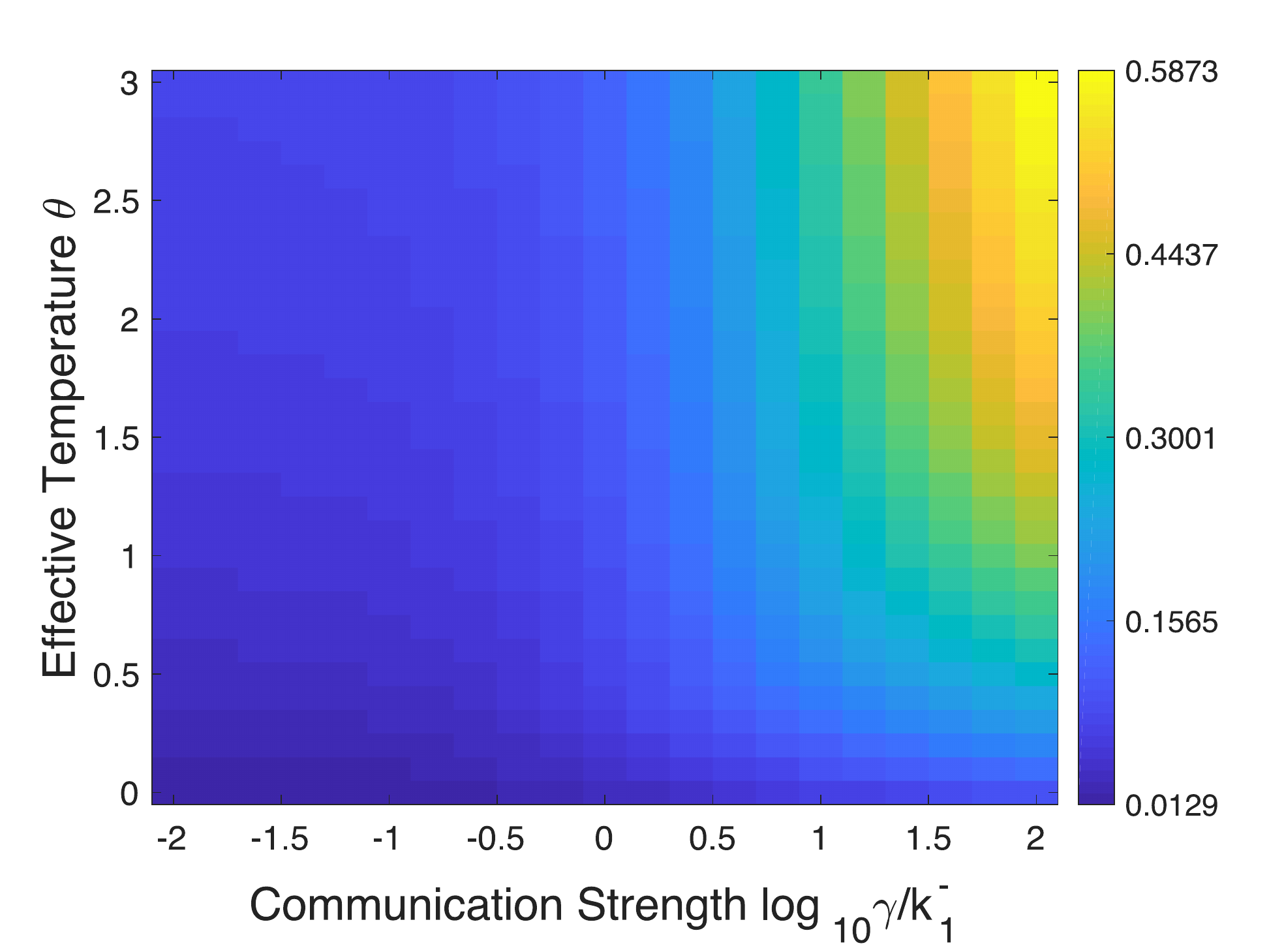}
	\caption{Information rate between a single cell's readout and the spatial average of the ligand fluctuations. This was obtained from the cross-spectrum under the linear noise approximation and numerically integrating. This plot has $N=10$, $h_i=0$, $x_c = 300$, $\overline{\ell} =150$, and $k_{\ell}^-/k_1^-=\gamma'/k_1^-=1$. The rate is maximized if we increase the effective temperature and communication strength.}
	\label{fig:rate}
\end{figure}

Finally, we again consider the information rate. Because the mutual information involving the spatial average was the only one to benefit from communication, we focus on this case. We compute the cross-spectrum under the linear noise approximation analytically (see Appendix F) and integrate over frequency numerically to find the information rate. The result is shown in Fig. \ref{fig:rate}. We see that the information rate is maximized when communication is strong and feedback is weak. Thus, as with the steady-state information, cell-cell communication improves the rate of information acquisition for this type of measure. However, like in the single-cell case, critical slowing down makes the rate suboptimal at criticality, such that the case without feedback has the highest rate. We also find that the rate increases with the ratio of sizes $x_c/\overline{\ell}$, decreases with the number of sites $N$ and the ratio of timescales $k_{\ell}^-/k_1^-$, and increases weakly with the ligand hopping rate $\gamma'$. Additionally, there can be a local maximum as $\gamma/k_1^-\rightarrow 0$ if $k_{\ell}^-/k_1^-$, but this isn't a global maximum, as the large $\gamma/k_1^-$ can exceed this value.

\section{Discussion}

We have investigated a minimal biochemical model with communication and intracellular feedback, in order to elucidate the tradeoffs of criticality for multicellular sensing. Criticality arises due to a bifurcation in the biochemical dynamics that places the model in the static universality class of the mean-field Ising model.
We have found that the susceptibility and noise both peak at the critical point. If the ligand fluctuations are sufficiently slow, the former effect dominates, and the mutual information is maximized. Otherwise, the information is maximized far from the critical point, where there is no feedback.
The introduction of cell-cell communication leads to long-range correlations. We have found that this feature leads to an increase in the information that a single cell gains about the average environment across the population, but not about the local or spatially resolved environment. Finally, we have found that although critical feedback can maximize steady-state information, it minimizes the information rate due to critical slowing down. 

How do our results compare to related theoretical work? Previous work has investigated the effect of receptor cooperativity on sensing an average ligand concentration using an Ising-like model \cite{skoge2011cooperativity}. There it was assumed that the ligand binding and unbinding dynamics are fast, and it was found that the signal-to-noise ratio is maximized when the receptors were independent. This result is consistent with our finding that the mutual information is maximized without feedback when the ligand is fast ($k_{\ell}^- \gtrsim k_1^-$). Other work has investigated the propagation of information in a two-dimensional spin system where one spin evolves via the telegraph process and the others have Glauber dynamics \cite{meijers2019infocrit}. There it was also found that the mutual information is maximized only when the driving is slow, consistent with our results. It was further found that the information rate is maximized at a finite driving timescale and a finite, supercritical temperature, whereas here we find that the information rate is maximized for an infinitely fast ligand timescale and infinite effective temperature (no feedback). It will be interesting to investigate if this apparent difference stems from the different structure or dimensionality of the two systems.

What are the implications of our results for particular biological systems? Our findings suggest that in systems with communication and feedback, such as the Bicoid-Hunchback system in fruit flies or quorum sensing in bacteria, sensory information is maximized at the critical point so long as input fluctuations are slow and sufficient time is available to combat critical slowing down. In the case of the Bicoid-Hunchback system in particular, previous experimental work has indeed suggested that criticality helps cells respond to small changes in the morphogen profile \cite{krotov2013morphogenesis}, whereas other work has argued that cells are instead in the bistable regime \cite{lopes2008spatial}. The auditory system is also a well-known sensory system that appears to benefit from being near criticality \cite{hudspeth2010cochlea}. There is also a form of cell-cell communication, as cells with different characteristic frequencies are coupled to enhance their response to general auditory signals \cite{stoop2016auditory}. These features are consistent with our findings, although it is important to note that frequency detection is a different task than the task of detecting fluctuations in a noisy, uniform ligand signal as we consider here.

Is criticality via intracellular feedback beneficial to cellular sensing overall? In light of our findings, it may be that this type of criticality is more detrimental for sensing a uniform concentration than it is beneficial, for several reasons. The type of information that criticality benefits concerns fluctuations about the mean, not the mean concentration itself, and the biological relevance of this task is unclear. Furthermore, we have found that high mutual information is only observed at sufficiently large molecule numbers and when there is a timescale separation of at least two orders of magnitude between the cell and its environment. Finally, the information rate, which better accounts for the fact that we are looking at a dynamical process, is minimized at criticality.

The approach that we have used is very general and can be applied to other systems that admit a Langevin description. There are many other biochemical models whose expansions near the bifurcation reduce to our model, and different normal forms can describe other classes of models. This fact can be exploited to probe more specific biochemical mechanisms and more general sensory measures. Nonetheless, our model is minimal and neglects features such as extrinsic noise, bursting, and cell-to-cell variability that are known to play an important role in biochemical signaling \cite{friedman2006linking, mugler2009spectral, shahrezaei2008colored, horsthemke1984noise, cotari2013cell}. In the future it will be important to expand our model to include these more general features.

\section*{Acknowledgments}

M.\ V.\ would like to thank Farshid Jafarpour and Terrence Edmonds for helpful discussions. M.\ V.\ and A.\ M. were supported by the Simons Foundation (376198) and the National Science Foundation (MCB-1936761). A.\ E.\ was supported by the National Science Foundation through the Center for the Physics of Biological Function (PHY-1734030) and by the National Institutes of Health (R01 GM082938).

\section*{Appendix A: Mutual Information for a Multivariate Gaussian}

 Our goal is to calculate the mutual information between the various combinations of the components of $\vec{\delta \ell}$ and $\vec{\delta x}$. In general, note that we have
\begin{equation}
I(X,Y) = H(X) + H(Y) -H(X,Y),
\end{equation} 
where $X$ and $Y$ are different variables (possibly sets of them) and $H$ is the Shannon entropy (in nats). The linearized stochastic differential equations describe an Ornstein-Uhlenbeck process, and the solution is a Gaussian random variable. If $X$ comes from a $D$-dimensional Gaussian distribution with covariance matrix $\mathcal{C}$, one may show that the entropy is
\begin{equation}
H(X) = \frac{D}{2} \log(2\pi e) + \frac{1}{2} \log(\det(\mathcal{C})).
\end{equation}
A key property that is useful in this analysis is the fact that taking a marginal of a multi-dimensional Gaussian yields another Gaussian and does not alter the covariances between the remaining variables.

Suppose that we have some combination of Gaussian variables $Z$ that are partitioned into two sets $X$ and $Y$, where $Z$ has $N$ components, $X$ has $N_x$ components, and $Y$ has $N_y = N-N_x$ components (any additional variables have been integrated out). Our covariance matrix may be decomposed in the following way
\begin{equation}
\mathcal{C} = \begin{bmatrix}
\mathcal{C}_{XX} & \mathcal{C}_{XY} \\
\mathcal{C}_{YX} & \mathcal{C}_{YY}
\end{bmatrix},
\end{equation}
where $\mathcal{C}_{YX}=\mathcal{C}_{XY}^T$. The entropies of the two subsets are
\begin{equation}
\begin{gathered}
H(X) = \frac{N_x}{2} \log(2\pi e) +\frac{1}{2} \log(\det(\mathcal{C}_{XX})), \\
H(Y) = \frac{N-N_x}{2} \log(2\pi e) +\frac{1}{2} \log(\det(\mathcal{C}_{YY})).
\end{gathered}
\end{equation}	 
The joint entropy will cancel off the constants, so the mutual information between $X$ and $Y$ is
\begin{equation}\label{eq:gaussianMI}
I(X,Y) = -\frac{1}{2} \log\left(\frac{\det(\mathcal{C})}{\det(\mathcal{C}_{XX}) \det(\mathcal{C}_{YY})}\right).
\end{equation}

\section*{Appendix B: Mutual Information for a Single Site}

Our linearized equations may be written in the form
\begin{equation}\label{eq:ouprocess}
d\vec{Y}_t = \mathbb{A} \vec{Y}_t dt + \mathbb{B} d\vec{W}_t, 
\end{equation}
where $\vec{Y}_t = [\delta x(t), \delta \ell(t)]^T$ describes the molecule numbers, $\vec{W}_t = [W_{x}(t), W_{\ell}(t)]^T$ describes the noise in production and degradation, and the matrices are
\begin{equation}
\mathbb{A} = \begin{bmatrix}
-c & k_1^+ \\
0 & -k_{\ell}^-
\end{bmatrix}, \qquad \mathbb{B} = \begin{bmatrix}
\sqrt{2d(x_c)} & 0 \\
0 & \sqrt{2 k_{\ell}^+}
\end{bmatrix}.
\end{equation}
If the initial condition is $\vec{Y}_0$, the general solution (for general matrices and dimensionalities) is
\begin{equation}\label{eq:ousolution}
\vec{Y}_t = e^{\mathbb{A}t} \vec{Y}_0 + \int_0^t e^{\mathbb{A}(t-s)}\mathbb{B} d\vec{W}_s.
\end{equation}
This may be proven using the substitution $\vec{Z}_t = e^{-\mathbb{A}t}\vec{Y}_t$ and using It{\^o}'s lemma. We are interested when the fluctuations about the steady state mean, so we take $\vec{Y}_0=0$. 

We need to evaluate the steady state covariances. Since the means are zero, it suffices to compute $\big\langle Y_t^{(i)} Y_t^{(j)} \big\rangle$ as $t\rightarrow \infty$. This can be done directly with our solution. Using the It{\^o} isometry, changing the integration variable to $t'=t-s$ and taking the aforementioned limit, we find
\begin{equation}\label{eq:ouCovar}
\big\langle Y^{(i)}Y^{(j)} \big\rangle = \int_{0}^{\infty} [e^{\mathbb{A}t}\mathbb{B} \mathbb{B}^T (e^{\mathbb{A}t})^T]_{i,j} dt.
\end{equation}
for general matrices and dimensionalities. The matrix exponential in our case is
\begin{equation}\label{eq:scMExp}
e^{\mathbb{A}t} = \begin{bmatrix}
e^{-ct} & -k_1^+ \dfrac{e^{-ct}-e^{-k_{\ell}^- t}}{c-k_{\ell}^-}\\
0 & e^{-k_{\ell}^- t}
\end{bmatrix}.
\end{equation}
The result when $c=k_{\ell}^-$ may be obtained by using L' H{\^o}pital's rule. The full covariance matrix is
\begin{equation}\label{eq:scCovar}
\mathcal{C} = 
\big\langle\vec{Y}\vec{Y}^T \big\rangle = \begin{bmatrix}
\dfrac{d(x_c)}{c} + \dfrac{(k_1^+)^2 \overline{\ell}}{c(c+k_{\ell}^-)} & \dfrac{k_1^+ \overline{\ell}}{c+k_{\ell}^-} \\
\dfrac{k_1^+ \overline{\ell}}{c+k_{\ell}^-} & \overline{\ell}
\end{bmatrix}.
\end{equation}
The diagonal terms are the variances, while the off-diagonal terms are the covariance between $\delta x$ and $\delta \ell$.

Before moving forward, it will be helpful to express things in terms of the ``Landau" parameters $x_c$, $h$, and $\theta$. Using the definitions in the main text, one can solve for the reaction rates
\begin{gather}
\begin{gathered}
k_1^+ = k_1^- \frac{x_c}{3\overline{\ell}} \frac{1 + 3\theta + 3h}{1+ \theta}, \\
 k_2^+ = \frac{k_1^-}{x_c (1+\theta)}, \quad k_2^- = \frac{k_1^-}{3 x_c^2 (1+\theta)}.
\end{gathered}
\end{gather}
Using these, we may also solve for $d(x_c)$ and $c$
\begin{equation}
d(x_c) = \frac{k_1^- x_c (4+3\theta)}{3(1+\theta)}, \qquad c= \frac{k_1^- \theta}{1+\theta}.
\end{equation}
Note that as $\theta\rightarrow \infty$, these quantities become what you would expect in the absence of feedback, provided that we replace $x_c$ by the appropriate mean $\overline{x} = k_1^+ \overline{\ell}/k_1^-$.

Now we compute the mutual information between the ligand and readout using Eqs. \ref{eq:gaussianMI}, \ref{eq:scCovar} with the result
\begin{equation}
I(\delta x,\delta \ell) = \frac{1}{2} \log\left( 1 + \frac{c (k_1^+)^2 \overline{\ell}}{d(c+ k_{\ell}^-)^2 + k_{\ell}^- (k_1^+)^2 \overline{\ell}} \right).
\end{equation}
Casting the rates into expressions of the Landau parameters and setting $h=0$ gives
\begin{widetext}
\begin{equation}
\label{eq:Ithetafull}
I(\delta x,\delta \ell) = \frac{1}{2} \log \left(1 + \frac{\theta (3\theta +1)^2 }{3(\overline{\ell}/x_c) (3\theta +4)[(k_{\ell}^-/k_1^-)(\theta +1) +\theta]^2 + (k_{\ell}^-/k_1^-) (\theta +1)(3\theta +1)^2 } \right).
\end{equation}
\end{widetext}
This is a complicated expression that vanishes as $\theta\rightarrow0$ unless $k_{\ell}^-/k_1^- \rightarrow 0$. If the ligand timescales are slow, then this simplifies to the expression given in the text.

\section*{Appendix C: Information Rate for a Single Site}

The information rate for Gaussian process where a scalar signal $s$ linearly drives a scalar response $x$ is
\begin{equation}\label{eq:infoRate}
R = -\frac{1}{4} \int_{-\infty}^{\infty} \log \left(1 - \frac{|S_{s,x}(\omega)|^2}{S_{s,s}(\omega) S_{x,x}(\omega)}\right) d\omega.
\end{equation}
where $S$'s are elements of the cross-spectrum $\mathbb{S}$, which satisfies
\begin{equation}\label{eq:XSpecFT}
\big\langle\hat{\vec{y}}(\omega) \hat{\vec{y}}^{\dagger}(\omega')\big\rangle = 4\pi^2 \mathbb{S}(\omega) \delta(\omega-\omega'),
\end{equation}
where we have combined the signal and response variables into a single vector as before \cite{tostevin2009trajectories,munakata2006dynamic}. We are using the convention that the Fourier transform is defined as
\begin{equation}
\hat{\vec{y}}(\omega) = \int_{-\infty}^{\infty} \vec{y}(t)e^{-i\omega t}dt.
\end{equation}
A vector of independent delta-correlated white noises $\vec{\eta}$ has a constant cross-spectrum
\begin{equation}
\big\langle\hat{\vec{\eta}}(\omega) \hat{\vec{\eta}}^{\dagger}(\omega')\big\rangle =2\pi \mathbb{I} \delta(\omega-\omega').
\end{equation}

We can solve for the cross-spectrum by taking the Fourier transform of Eq. \ref{eq:scSDEs}.
The ligand doesn't depend on the readout, so we can solve for this first. The equation becomes
\begin{equation}
\hat{\delta \ell}(\omega) = \frac{\sqrt{2 k_{\ell}^+} \hat{\epsilon}(\omega)}{k_{\ell}^- +i\omega}.
\end{equation}
Using this to find the power spectrum of the ligand fluctuations yields
\begin{equation}
S_{\ell,\ell}(\omega)= \frac{k_{\ell}^+ }{\pi ((k_{\ell}^-)^2 +\omega^2)}.
\end{equation}
Taking the Fourier transform of the second equation and using the result from the first gives
\begin{equation}
\hat{\delta x}(\omega) = \frac{\sqrt{2d} \hat{\eta}(\omega)}{c+i\omega} + \frac{k_1^+}{c+i\omega}\frac{\sqrt{2 k_{\ell}^+} \hat{\epsilon}(\omega)}{k_{\ell}^- +i\omega}
\end{equation}
When finding the cross-spectrum between $\delta x$ and $\delta \ell$, the calculation is the same as the previous case, as the $\hat{\eta}$ term cancels, except we have an additional factor
\begin{equation}
S_{x,\ell}(\omega)= \frac{k_1^+}{c+i\omega} \frac{k_{\ell}^+ }{\pi ((k_{\ell}^-)^2 +\omega^2)}.
\end{equation}
For the power spectrum of $\delta x$, the two terms are independent and the result is
\begin{equation}
S_{x,x}(\omega) = \frac{d}{\pi(c^2 + \omega^2)} +\frac{(k_1^+)^2}{c^2 + \omega^2} \frac{k_{\ell}^+}{\pi((k_{\ell}^-)^2 +\omega^2)}.
\end{equation}
Plugging everything in, the information rate integral evaluates to
\begin{equation}
R = \frac{\pi}{2} \left( \sqrt{(k_{\ell}^-)^2 + \frac{(k_1^+)^2 k_{\ell}^+}{d}} - k_{\ell}^- \right).
\end{equation}
Expressing this in terms of the Landau parameters gives the result in the main text.

\section*{Appendix D: Covariance Matrix for Multiple Cells}

Our system of stochastic differential equations may be expressed in terms of circulant matrices. These matrices have a number of nice properties that will be used in the analysis, so it will be worthwhile to discuss them. A circulant matrix $\mathbb{M}$ is a square matrix such that the next column can be obtained by shifting the entries of the current column down by one and imposing periodicity. Concretely, they take the form
\begin{equation}
\mathbb{M}=
\begin{bmatrix}
m_0     & m_{N-1} & \dots  & m_{2} & m_{1}  \\
m_{1} & m_0    & m_{N-1} &         & m_{2}  \\
\vdots  & m_{1}& m_0    & \ddots  & \vdots   \\
m_{N-2}  &        & \ddots & \ddots  & m_{N-1}   \\
m_{N-1}  & m_{N-2} & \dots  & m_{1} & m_0 \\
\end{bmatrix}.
\end{equation}
The discrete Laplacian on a ring with $N$ sites is a special case of this with $m_0=-2$, $m_1=m_{N-1}=1$, and zeroes in all other entries. All circulant matrices are simultaneously diagonalizable via a discrete Fourier transform. Since they are simultaneously diagonalizable, they all commute with each other. The set of circulant matrices is closed under matrix addition and multiplication. If a circulant matrix is invertible, its inverse is another circulant matrix. For each $j\in \{0,1,...,N-1\}$, there is an eigenvector
\begin{equation}
	\vec{v}_j = \frac{1}{\sqrt{N}} \begin{bmatrix}
	1 & e^{-2\pi i j/N} & ... & e^{-2\pi i j(N-1)/N}
	\end{bmatrix}^T,
\end{equation}
with eigenvalue
\begin{equation}
\lambda_{\mathbb{M}}(j) = \sum_{k=0}^{N-1} m_k e^{2\pi i jk/N}.
\end{equation}
It is easy to check that this eigenbasis is orthonormal. For the discrete Laplacian, the eigenvalues are
\begin{equation}
	\lambda_{\nabla^2}(j) = -2[1-\cos(2\pi j/N)].
\end{equation}
We will use $U$ to denote the unitary discrete Fourier transform, i.e. the matrix whose columns are the eigenvectors ordered from $j=0,1,...,N-1$.

Now we will solve the system by casting it into the canonical form for an Ornstein-Uhlenbeck process and then using Eq. \ref{eq:ousolution}. As before, we introduce a vector $\vec{Y}_t = [\vec{\delta x}(t), \vec{\delta \ell}(t)]^T$ that describes both molecular profiles. We introduce a $6N$-dimensional vector of noises that are ordered as follows
	\begin{equation}
d\vec{W}_t = \begin{bmatrix}
\text{X chemical noises} \\
\text{X diffuse left noises} \\
\text{X diffuse right noises} \\
\text{L chemical noises} \\
\text{L diffuse left noises} \\
\text{L diffuse right noises}
\end{bmatrix}.
\end{equation}
The matrices $\mathbb{A}$ and $\mathbb{B}$ are best expressed in block form. If $\mathbb{I}_N$ is the $N$-dimensional identity matrix, we have
\begin{equation}
	\mathbb{A} = \begin{bmatrix}
	-c \mathbb{I}_N + \gamma \nabla^2 & k_1^+ \mathbb{I}_N \\
	0 & -k_{\ell}^- \mathbb{I}_N + \gamma' \nabla^2
	\end{bmatrix}.
\end{equation}
The $\mathbb{B}$ matrix takes the form
\begin{widetext}
\begin{equation}
\mathbb{B} = \begin{bmatrix}
\sqrt{2 d(x_c)} \mathbb{I}_N & \sqrt{\gamma x_c} D_L & \sqrt{\gamma x_c} D_R & 0 & 0 & 0 \\
0 & 0 & 0 & \sqrt{2 k_{\ell}^+} \mathbb{I}_N  & \sqrt{\gamma' \overline{\ell}} D_L & \sqrt{\gamma' \overline{\ell}} D_R
\end{bmatrix},
\end{equation}
\end{widetext}
where $D_L$ ($D_R$) is a circulant matrix with $m_0=-1$, $m_{N-1}=1$ ($m_1=1$), and zero for all of the other entries. These matrices describe the anti-correlations associated with diffusing left and right respectively and they satisfy
\begin{equation}
\begin{gathered}
D_L^T = D_R, \quad D_L D_R =D_R D_L = -\nabla^2.
\end{gathered}
\end{equation}
With these definitions, our system of stochastic differential equations is in the standard form and the usual solution applies.

Next, we need to evaluate the matrices and integration that appear in Eq. \ref{eq:ouCovar}. The product of $\mathbb{B}$ matrices is easy
\begin{equation}
	\mathbb{B}\mathbb{B}^T =  \begin{bmatrix}
	2d(x_c) \mathbb{I}_N -2\gamma x_c \nabla^2 & 0 \\
	0 & 2k_{\ell}^+ \mathbb{I}_N -2\gamma' \overline{\ell} \nabla^2
	\end{bmatrix}
\end{equation}
Computing the matrix exponential is more involved. It is instructive to work with a $2\times 2$ matrix. Consider the family of matrices
\begin{equation}
\mathbb{M}_N = \begin{bmatrix}
a\mathbb{I}_N & b\mathbb{I}_N \\
0 & c\mathbb{I}_N
\end{bmatrix}.
\end{equation}
For $N=1$, the matrix exponential is 
\begin{equation}
e^{\mathbb{M}_1 t} = \begin{bmatrix}
e^{at} & b\dfrac{e^{at} - e^{c t}}{a-c} \\
0 & e^{ct}
\end{bmatrix}.
\end{equation}
Working with the number $1$ is similar to working with the identity matrix, both have multiplicative inverses and have commutative multiplication. In fact, a similar solution holds for arbitrary $N$
\begin{equation}
e^{\mathbb{M}_N t} = \begin{bmatrix}
e^{a\mathbb{I}_N t} & b(a\mathbb{I}_N -c\mathbb{I}_N)^{-1}(e^{a \mathbb{I}_N t} - e^{c \mathbb{I}_N t}) \\
0 & e^{c\mathbb{I}_N t}
\end{bmatrix},
\end{equation}
here we have written it in a suggestive form. For our problem, it will be convenient to introduce the shorthand
\begin{equation}
\begin{gathered}
\alpha = -c\mathbb{I}_N + \gamma \nabla^2, \\
\beta = -k_{\ell}^-\mathbb{I}_N + \gamma' \nabla^2, \\
\alpha' = 2d(x_c) \mathbb{I}_N -2\gamma x_c \nabla^2, \\
\beta' = 2k_{\ell}^+ \mathbb{I}_N -2\gamma' \overline{\ell} \nabla^2 = -2\overline{\ell} \beta.
\end{gathered}
\end{equation}
All of these matrices are circulant, so they commute and are diagonalizable. The first two have strictly negative eigenvalues, while the last two have positive eigenvalues, so they are all invertible. In light of the single cell result in Eq. \ref{eq:scMExp}, a natural candidate for the exponential of $\mathbb{A}$ is
\begin{equation}
	e^{\mathbb{A} t} = \begin{bmatrix}
	e^{\alpha t} & -k_1^+ (\alpha-\beta)^{-1} (e^{\alpha t} - e^{\beta t}) \\
	0 & e^{\beta t}
	\end{bmatrix}.
\end{equation}
The matrix exponential $\mathbb{E}(t)=e^{\mathbb{A} t}$ is the unique solution to the initial value problem
\begin{equation}
	\frac{d \mathbb{E}(t)}{dt} = \mathbb{A}\mathbb{E}(t),\quad \mathbb{E}(0) = \mathbb{I}_{2N}.
\end{equation}
Using the fact that all of the matrices involved commute, one may show that our guess satisfies these equations. All that remains is evaluating the product in the integrand and computing the integral. This is greatly facilitated by the fact that all matrices involved commute and are invertible. The final result for the steady state covariance matrix is
\begin{widetext}
\begin{equation}\label{eq:mcCovar}
\mathcal{C} = \begin{bmatrix}
-\dfrac{1}{2} \alpha^{-1} [\alpha' -2 (k_1^+)^2 \overline{\ell} (\alpha+\beta)^{-1}] & -k_1^+\overline{\ell} (\alpha+\beta)^{-1} \\
-k_1^+\overline{\ell} (\alpha+\beta)^{-1} & \overline{\ell} \mathbb{I}_N
\end{bmatrix}.
\end{equation}
\end{widetext}
Note that $\mathcal{C}_{\vec{\delta x},\vec{\delta \ell}} = \mathcal{C}_{\vec{\delta \ell},\vec{\delta x}}$ = $\mathcal{C}_{\vec{\delta x},\vec{\delta \ell}}^T$, this is a manifestation of translational invariance. With the full covariance matrix, we may compute the steady state mutual information between any pair of combinations of the variables.

\section*{Appendix E: Mutual Information for Multiple Sites}

Now that we are interested in a single cell's readout, we can start reducing the covariance matrix. We only want to track a single cell, say the cell at site $0$, so we only need one diagonal entry from $\mathcal{C}_{\vec{\delta x},\vec{\delta x}}$. If $\mathbb{M}$ is a matrix with a constant diagonal, then
\begin{equation}
\mathbb{M}_{0,0} = \frac{1}{N} \sum\limits_{j=0}^{N-1} \lambda_{\mathbb{M}}(j).
\end{equation}
If we assume that the ligand timescales are slow, we have
\begin{equation}
\sigma_x^2 = -\frac{1}{2N} \sum_{j=0}^{N-1} \left[ \frac{\lambda_{\alpha'}(j)}{\lambda_{\alpha}(j)} - \frac{2 (k_1^+)^2 \overline{\ell}}{\lambda_{\alpha}(j)^2}\right],
\end{equation}
where the eigenvalues are 
\begin{equation}
\begin{gathered}
\lambda_{\alpha}(j) = -c -2\gamma \left[ 1-\cos(2\pi j/N) \right], \\
\lambda_{\alpha'}(j) = 2d(x_c) +4\gamma x_c \left[ 1-\cos(2\pi j/N) \right].
\end{gathered}
\end{equation}
The first term describes the intrinsic noise, while the second describes the extrinsic noise. We can evaluate the sum approximately in the large $N$ limit by adding and subtracting a $j=N$ term and using the Euler-Maclaurin formula \cite{bender1999asymptotics}. The formula states that, for a smooth $f$, we have
\begin{equation}
\begin{aligned}
	\sum\limits_{j=0}^{n} &f(j) = \int_{0}^{n}f(t)dt + \frac{f(n) + f(0)}{2} \\
	&+ \sum\limits_{j=1}^{k} \frac{(-1)^{j+1}B_{j+1}}{(j+1)!} [f^{(j)}(n) - f^{(j)}(0)] + R_k,
\end{aligned}
\end{equation}
where the remainder term is
\begin{equation}
	R_k = \frac{(-1)^k}{(k+1)!} \int_0^n f^{(k+1)}(t) P_{k+1}(t)dt,
\end{equation}
the Bernoulli numbers are denoted by the $B_k$, and $P_k(t) = B_k(t-\lfloor t \rfloor )$ denotes the periodic Bernoulli functions. With the sum extended to $N$, there is tremendous simplification, as $f^{(k)}(N) = f^{(k)}(0)$. We had to add and subtract the $j=N$ term; the added term gets taken into the integral, while the subtracted term is canceled by the average of the two endpoints. To evaluate the integrals, it is best to work with the angular variables $\phi=2\pi t/N$, which ranges over $[0,2\pi]$. The remainder term becomes smaller as $N$ increases; the terms in the Fourier series for the periodic Bernoulli functions have arguments of the form $mN\phi$, where $m$ is a non-zero integer. Neglecting the remainder term, we find
\begin{equation}
\begin{split}
	\sigma_x^2 = \left(x_c + \frac{d(x_c) -c x_c}{\sqrt{c(c+4\gamma)}}\right) + \frac{(k_1^+)^2 \overline{\ell} (c+2\gamma)}{(c(c+4\gamma))^{3/2}},
\end{split}
\end{equation}
where the first pair describes intrinsic noise and the last term describes extrinsic noise. Generally speaking, this continuum approximation will break down on some neighborhood of the critical point. For example, the sum for the variance diverges like $\theta^{-2}$ as we approach the critical point, but the result of the integral diverges like $\theta^{-3/2}$. Nonetheless, the two results agree well numerically on all but a small neighborhood of the critical point, and the results still diverge, so we expect it to give reasonably accurate results.

We will start with the on-site mutual information. The variance of each ligand molecules is $\overline{\ell}$, so it suffices to compute the covariance between the ligand and readout. In the slow ligand and continuum limits, this is
\begin{equation}
\big\langle (\delta x_0)(\delta \ell_0) \big\rangle =  \frac{k_1^+ \overline{\ell}}{\sqrt{c(c+4\gamma)}}.
\end{equation}
For a pair of Gaussian variables, the mutual information increases monotonically with the ratio $\langle (\delta x)(\delta y)\rangle^2/\sigma_x^2 \sigma_y^2$. When expressed in terms of the Landau parameters and using $\theta > 0$, the derivative with respect to $\gamma$ is negative. This means that the mutual information decreases with increasing communication strength. 

We will compute the result for the spatially resolved profile. We need to reduce the covariance matrix. We do this by integrating out all but one of the readout molecules. This amounts to working with the reduced covariance matrix
\begin{equation}
	\mathcal{C}_{\text{red}} = \begin{bmatrix}
	\sigma_x^2 & \hat{e}^T_0 \mathcal{C}_{\vec{\delta x},\vec{\delta \ell}} \\
	\mathcal{C}_{\vec{\delta x},\vec{\delta \ell}} \hat{e}_0 & \overline{\ell} \mathbb{I}_N 
	\end{bmatrix}.
\end{equation}
The unit vector $\hat{e}_0 = [1, 0, ..., 0]^T$ is used to extract the first row and column from the matrix describing the correlations between $x$ and $\ell$. In order to move forward, we use a result about the determinants of block matrices. Suppose that $\mathbb{M}$ takes the form
\begin{equation}
\mathbb{M} = \begin{bmatrix}
A & B \\
C & D
\end{bmatrix},
\end{equation}
where $A$ and $D$ are square matrices of potentially different sizes and $D$ is invertible. Then we have
\begin{equation}
\det(\mathbb{M}) = \det(A - BD^{-1}C) \det(D).
\end{equation}
It follows that the determinant in the numerator of the mutual information is
\begin{equation}
	\det(\mathcal{C}_{\text{red}}) = (\sigma_x^2 - \hat{e}_0^T \mathcal{C}_{\vec{\delta x},\vec{\delta \ell}}^2 \hat{e}_0/\overline{\ell})\overline{\ell}^N
\end{equation}
and the mutual information is
\begin{equation}
I(\delta x_0,\vec{\delta \ell}) = -\frac{1}{2} \log\left(1 - \frac{\hat{e}_0^T \mathcal{C}_{\vec{\delta x},\vec{\delta \ell}}^2 \hat{e}_0}{\overline{\ell} \sigma_x^2}\right).
\end{equation}
The $\mathcal{C}_{\vec{\delta x},\vec{\delta \ell}}$ appearing in the numerator is circulant, as it is the inverse of a circulant matrix, so its square is also circulant. We may apply the formula derived at the beginning of the section in the slow ligand limit to get
\begin{equation}
\hat{e}_0^T \mathcal{C}_{\vec{\delta x},\vec{\delta \ell}}^2 \hat{e}_0 = \frac{1}{N} \sum_{j=0}^{N-1} \left(\frac{k_1^+ \overline{\ell}}{\lambda_{\alpha}(j)}\right)^2.
\end{equation}
This is the sum that we encountered when computing the extrinsic noise multiplied by $\overline{\ell}$. It follows that the mutual information may be written in the form
\begin{equation}
I(\delta x_0,\vec{\delta \ell}) = \frac{1}{2} \log\left(1+ \frac{\sigma_{\text{ext}}^2}{\sigma_{\text{int}}^2}\right).
\end{equation}
The mutual information increases monotonically with the ratio of contributions to the noise. Expressing it in terms of $\theta$ and differentiating with respect to $\gamma$, the result is negative, so communication also impairs this information.

We finally turn to calculating the mutual information between a single readout and the spatial average of the ligand fluctuations. Since the ligand molecules at each site are identically and independently distributed Poisson variables with mean $\overline{\ell}$, the variance in the spatial average of the ligand is
\begin{equation}
\sigma_L^2 = \frac{\overline{\ell}}{N}.
\end{equation}
We need to compute the covariance between a single cell and this average. This can be found by averaging the first row or column of the readout-ligand covariance matrix
\begin{equation}
\big\langle (\delta x_0)(\delta L)\big\rangle = \frac{1}{N} \hat{e}^T_0 \mathcal{C}_{\vec{\delta x},\vec{\delta \ell}} \vec{1}_N,
\end{equation}
where $\vec{1}_N$ is a $N$-dimensional vector whose components are all $1$. This is an eigenvector of circulant matrices, so we have
\begin{equation}
U^{\dagger} \vec{1} = \sqrt{N} \hat{e}_0.
\end{equation}
Using the unitary discrete Fourier transform to diagonalize the covariance matrix, we find that this picks off the eigenvalue with $j=0$ that does not depend on diffusion
\begin{equation}
\big\langle (\delta x_0)(\delta L)\big\rangle = \frac{k_1^+ \overline{\ell}}{N c}.
\end{equation}
Squaring and dividing by the variance in $\delta L$, we see that this is the $j=0$ term in the extrinsic noise. If we work with the variance as a sum of eigenvalues, the ratio of the squared covariance to the products of the noises tends to one as we approach the critical point, so the mutual information diverges there. If we use the integral approximation to the variance, this ratio can exceed one, leading to a negative argument. This approximation breaks down for small $\theta$, but we can trust it away from a small neighborhood of the critical point. Differentiating the ratio that appears in the mutual information with respect to $\gamma$ gives a positive answer, so communication helps this mutual information.

\section*{Appendix F: Information Rate for Multiple Sites}

Since the mutual information between a single cell and the spatial average of the ligand fluctuations was the only case that benefited from communication, we will restrict our focus to this case. We will take a different approach than we did for a single site, as the Langevin equations are now a $2N$-dimensional coupled linear system. Using the same convention with Fourier transforms as before, the Wiener-Khinchin theorem states that
\begin{equation}
	\mathbb{S}(\omega) = \frac{1}{2\pi} \int_{-\infty}^{\infty} \mathcal{C}(\tau) e^{-i\omega \tau} d\tau,
\end{equation}
where $\mathcal{C}(\tau)$ is the steady state correlation matrix. To capture the steady state correlations, we initialize the system at $t_0=-\infty$. Since our variables have zero mean, this takes the form
\begin{equation}
\mathcal{C}(\tau) = \big\langle \vec{Y}_{t+\tau} \vec{Y}_t^T \big\rangle.
\end{equation}
Since we initialized this at $t_0 = -\infty$, this doesn't depend on the choice of $t$. When $\tau=0$, this reduces to the covariance matrix $\mathcal{C}$ that we worked with before. Using the It{\^o} isometry and performing a change of variables to $t'=t-s$, one may show that
\begin{equation}
	\mathcal{C}(\tau) = \int_{-\min(0,\tau)}^{\infty} e^{\mathbb{A}(t+\tau)} \mathbb{B} \mathbb{B}^T (e^{\mathbb{A} t})^T dt.
\end{equation}
When $\tau \geq 0$, we may factor off $e^{\mathbb{A}\tau}$ to the left, and the remaining integral becomes $\mathcal{C}$. For $\tau<0$, we perform a change of variables to $t'=t+\tau$, then factoring off $e^{-\mathbb{A}^T \tau}$ to the right gives the integral that yields $\mathcal{C}$. In summary, we have
\begin{align}
	\mathcal{C}(\tau) = \begin{cases}
	e^{\mathbb{A} \tau} \mathcal{C} , \quad &\tau\geq 0, \\
	\mathcal{C} (e^{-\mathbb{A} \tau})^T, \quad &\tau<0.
	\end{cases}
\end{align}
Using this, we find that the cross-spectrum is
\begin{equation}
	\mathbb{S}(\omega) = -\frac{\left[ (\mathbb{A} -i\omega \mathbb{I}_{2N})^{-1} \mathcal{C} + \mathcal{C}(\mathbb{A}^T +i\omega \mathbb{I}_{2N})^{-1}\right]}{2\pi} .
\end{equation}
To evaluate this, we need to find the inverse matrix and then plug it in. Since the $\alpha$, $\beta$, and $\mathbb{I}_N$ all commute and are invertible, treating them as if they were scalars leads to the guess
\begin{widetext}
\begin{equation}
(\mathbb{A}-i\omega \mathbb{I}_{2N})^{-1} = \begin{bmatrix}
(\alpha -i\omega \mathbb{I}_N)^{-1} & -k_1^+ (\alpha -i\omega \mathbb{I}_N)^{-1}(\beta -i\omega \mathbb{I}_N)^{-1} \\
0 & (\beta -i\omega \mathbb{I}_N)^{-1}
\end{bmatrix},
\end{equation}
\end{widetext}
and a direct computation shows that this is the inverse. Using this, we can compute the cross-spectrum
\begin{widetext}
\begin{equation}
\mathbb{S}(\omega) = -\frac{1}{2\pi} \begin{bmatrix}
 (\alpha^2 +\omega^2 \mathbb{I}_N)^{-1}[ -\alpha' + 2(k_1^+)^2 \overline{\ell} \beta (\beta^2 +\omega^2 \mathbb{I}_N)^{-1}] & -2 k_1^+ \overline{\ell} \beta (\beta^2 +\omega^2 \mathbb{I}_N)^{-1} (\alpha- i\omega \mathbb{I}_N)^{-1} \\
 -2 k_1^+ \overline{\ell} \beta (\beta^2 +\omega^2 \mathbb{I}_N)^{-1} (\alpha+ i\omega \mathbb{I}_N)^{-1} & 2\overline{\ell} \beta (\beta^2 +\omega^2 \mathbb{I}_N)^{-1}
\end{bmatrix}.
\end{equation}
\end{widetext}
Now we need to compute the relevant terms from these matrices.

First, we compute the power spectrum for the fluctuations in $\delta x_0$. Since all of the matrices involved here are circulant, any diagonal element is the average of the eigenvalues
\begin{equation}
\begin{aligned}
S_{\delta x_0,\delta x_0}(\omega) &= -\frac{1}{2\pi N} \sum_{j=0}^{N-1} \left[ \frac{-\lambda_{\alpha'}(j)}{\lambda_{\alpha}(j)^2 +\omega^2}  \right. \\
&\left.+ \frac{2(k_1^+)^2 \overline{\ell} \lambda_{\beta}(j)}{(\lambda_{\alpha}(j)^2 +\omega^2)(\lambda_{\beta}(j)^2 +\omega^2)}\right],
\end{aligned}	
\end{equation}
where the eigenvalues of $\beta$ are
\begin{equation}
\lambda_{\beta}(j) = -k_{\ell}^- -2\gamma' (1-\cos(2\pi j/N)).
\end{equation}
It is possible to approximate this as an integral and evaluate the integral analytically. However, the result is complicated and appears in an integral as part of a logarithm's argument. We cannot make progress with the final integral, so we will evaluate this numerically.

One may show that the power spectrum in $\delta L$ is the average of all of the matrix elements in the cross-spectrum $\mathbb{S}_{\vec{\delta \ell},\vec{\delta \ell}}$ by using the definition of $\delta L$ and Eq. \ref{eq:XSpecFT}. We can write this compactly using the vector of ones
\begin{equation}
	S_{\delta L, \delta L}(\omega) = -\frac{1}{2\pi N^2} \vec{1}^T_N \mathbb{S}_{\vec{\delta \ell},\vec{\delta \ell}} \vec{1}_N.
\end{equation}
Since $\mathbb{S}_{\vec{\delta \ell},\vec{\delta \ell}}$ is circulant, the vector of ones is an eigenvector of the matrix with $j=0$, so we have
\begin{equation}
S_{\delta L, \delta L}(\omega) = \frac{\overline{\ell} k_{\ell}^-}{\pi N[(k_{\ell}^-)^2 +\omega^2]}.
\end{equation}

Finally, we compute the cross-spectrum between $\delta x_0$ and $\delta L$. This should be the average of the elements in the first row in $\mathbb{S}_{\vec{\delta x},\vec{\delta \ell}}$, which may be obtained via
\begin{equation}
	S_{\delta x_0,\delta L}(\omega) = -\frac{1}{2\pi N} \hat{e}_0^T \mathbb{S}_{\vec{\delta x},\vec{\delta \ell}} \vec{1}_N.
\end{equation}
The vector of ones is also an eigenvector of this matrix with $j=0$, so
\begin{equation}
S_{\delta x_0,\delta L}(\omega) = \frac{k_1^+ k_{\ell}^- \overline{\ell}}{\pi N[(k_{\ell}^-)^2 +\omega^2][c+i\omega]}.
\end{equation}
With the power spectra and cross-spectrum, we can numerically integrate Eq. \ref{eq:infoRate}. A detailed description of how this depends on all of the parameters is provided in the main text.


\begin{thebibliography}{48}
\expandafter\ifx\csname natexlab\endcsname\relax\def\natexlab#1{#1}\fi
\expandafter\ifx\csname bibnamefont\endcsname\relax
  \def\bibnamefont#1{#1}\fi
\expandafter\ifx\csname bibfnamefont\endcsname\relax
  \def\bibfnamefont#1{#1}\fi
\expandafter\ifx\csname citenamefont\endcsname\relax
  \def\citenamefont#1{#1}\fi
\expandafter\ifx\csname url\endcsname\relax
  \def\url#1{\texttt{#1}}\fi
\expandafter\ifx\csname urlprefix\endcsname\relax\def\urlprefix{URL }\fi
\providecommand{\bibinfo}[2]{#2}
\providecommand{\eprint}[2][]{\url{#2}}

\bibitem{berg1977chemoreception}
H~C Berg and E~M Purcell.
\newblock Physics of chemoreception.
\newblock {\em Biophysical Journal}, 20(2):193--219, November 1977.

\bibitem{kaizu2014revisited}
Kazunari Kaizu, Wiet de Ronde, Joris Paijmans, Koichi Takahashi, Filipe
Tostevin, and Pieter Rein ten Wolde.
\newblock The {Berg}-{Purcell} {Limit} {Revisited}.
\newblock {\em Biophysical Journal}, 106(4):976--985, February 2014.

\bibitem{ngampruetikorn2020energy}
Vudtiwat Ngampruetikorn, David~J. Schwab, and Greg~J. Stephens.
\newblock Energy consumption and cooperation for optimal sensing.
\newblock {\em Nature Communications}, 11(975):1--8, February 2020.

\bibitem{carballopacheco2019xtalk}
Martín Carballo-Pacheco, Jonathan Desponds, Tatyana Gavrilchenko, Andreas
Mayer, Roshan Prizak, Gautam Reddy, Ilya Nemenman, and Thierry Mora.
\newblock Receptor crosstalk improves concentration sensing of multiple
ligands.
\newblock {\em Physical Review E}, 99(2):022423, February 2019.

\bibitem{mora2015spurious}
Thierry Mora.
\newblock Physical {Limit} to {Concentration} {Sensing} {Amid} {Spurious}
{Ligands}.
\newblock {\em Physical Review Letters}, 115(3):038102, July 2015.

\bibitem{endres2008gradient}
Robert~G. Endres and Ned~S. Wingreen.
\newblock Accuracy of direct gradient sensing by single cells.
\newblock {\em Proceedings of the National Academy of Sciences},
105(41):15749--15754, October 2008.

\bibitem{hu2010gradients}
Bo~Hu, Wen Chen, Wouter-Jan Rappel, and Herbert Levine.
\newblock Physical {Limits} on {Cellular} {Sensing} of {Spatial} {Gradients}.
\newblock {\em Physical Review Letters}, 105(4):048104, July 2010.

\bibitem{gregor2007positional}
Thomas Gregor, David~W. Tank, Eric~F. Wieschaus, and William Bialek.
\newblock Probing the {Limits} to {Positional} {Information}.
\newblock {\em Cell}, 130(1):153--164, July 2007.

\bibitem{ellison2016communication}
David Ellison, Andrew Mugler, Matthew~D. Brennan, Sung~Hoon Lee, Robert~J.
Huebner, Eliah~R. Shamir, Laura~A. Woo, Joseph Kim, Patrick Amar, Ilya
Nemenman, Andrew~J. Ewald, and Andre Levchenko.
\newblock Cell–cell communication enhances the capacity of cell ensembles to
sense shallow gradients during morphogenesis.
\newblock {\em Proceedings of the National Academy of Sciences},
113(6):E679--E688, February 2016.

\bibitem{fancher2017collective}
Sean Fancher and Andrew Mugler.
\newblock Fundamental {Limits} to {Collective} {Concentration} {Sensing} in
{Cell} {Populations}.
\newblock {\em Physical Review Letters}, 118(7):078101, February 2017.

\bibitem{treisman1989products}
Jessica Treisman and Claude Desplan.
\newblock The products of the {\textit{drosophila}} gap genes
{\textit{hunchback}} and {\textit{krüppel}} bind to the {\textit{hunchback}}
promoters.
\newblock {\em Nature}, 341(6240):335--337, September 1989.

\bibitem{erdmann2009averaging}
Thorsten Erdmann, Martin Howard, and Pieter~Rein ten Wolde.
\newblock Role of {Spatial} {Averaging} in the {Precision} of {Gene}
{Expression} {Patterns}.
\newblock {\em Physical Review Letters}, 103(25):258101, December 2009.

\bibitem{williams2008quorum}
Joshua~W Williams, Xiaohui Cui, Andre Levchenko, and Ann~M Stevens.
\newblock Robust and sensitive control of a quorum-sensing circuit by two
interlocked feedback loops.
\newblock {\em Molecular Systems Biology}, 4(1):234, January 2008.

\bibitem{strogatz2018nonlinear}
Steven~H. Strogatz.
\newblock {\em Nonlinear {Dynamics} and {Chaos}: {With} {Applications} to
	{Physics}, {Biology}, {Chemistry}, and {Engineering}}.
\newblock CRC Press, May 2018.

\bibitem{erez2019universality}
Amir Erez, Tommy~A. Byrd, Robert~M. Vogel, Grégoire Altan-Bonnet, and Andrew
Mugler.
\newblock Universality of biochemical feedback and its application to immune
cells.
\newblock {\em Physical Review E}, 99(2):022422, February 2019.

\bibitem{byrd2019slow}
Tommy~A. Byrd, Amir Erez, Robert~M. Vogel, Curtis Peterson, Michael
Vennettilli, Grégoire Altan-Bonnet, and Andrew Mugler.
\newblock Critical slowing down in biochemical networks with feedback.
\newblock {\em Physical Review E}, 100(2):022415, August 2019.

\bibitem{mora2011criticality}
Thierry Mora and William Bialek.
\newblock Are {Biological} {Systems} {Poised} at {Criticality}?
\newblock {\em Journal of Statistical Physics}, 144(2):268--302, July 2011.

\bibitem{krotov2013morphogenesis}
Dmitry Krotov, Julien~O. Dubuis, Thomas Gregor, and William Bialek.
\newblock Morphogenesis at criticality.
\newblock {\em Proceedings of the National Academy of Sciences},
111(10):3683--3688, March 2014.

\bibitem{munoz2018criticality}
Miguel~A. Muñoz.
\newblock Colloquium: {Criticality} and dynamical scaling in living systems.
\newblock {\em Reviews of Modern Physics}, 90(3):031001, July 2018.

\bibitem{schlogl1972chemical}
F.~Schlögl.
\newblock Chemical reaction models for non-equilibrium phase transitions.
\newblock {\em Zeitschrift für Physik}, 253(2):147--161, April 1972.

\bibitem{bose2019criticality}
Indrani Bose and Sayantari Ghosh.
\newblock Bifurcation and criticality.
\newblock {\em Journal of Statistical Mechanics: Theory and Experiment},
2019(4):043403, April 2019.

\bibitem{goldenfeld1992renorm}
Nigel Goldenfeld.
\newblock {\em Lectures {On} {Phase} {Transitions} {And} {The}
	{Renormalization} {Group}}.
\newblock Addison-Wesley, Reading, Mass, June 1972.

\bibitem{erez2019twocell}
Amir Erez, Tommy~A. Byrd, Michael Vennettilli, and Andrew Mugler.
\newblock Cell-to-cell information at a feedback-induced bifurcation point.
\newblock {\em Physical Review Letters}, 125:048103, July 2020.

\bibitem{zhang2010vanthoff}
Yunxin Zhang, Hao Ge, and Hong Qian
\newblock van't Hoff-Arrhenius Analysis of Mesoscopic and Macroscopic Dynamics of Simple Biochemical Systems: Stochastic vs. Nonlinear Bistabilities.
\newblock {\em 	arXiv:1011.2554}, November 2010.

\bibitem{brachet1981schlogl2}
M.~E. Brachet and E.~Tirapegui.
\newblock On the critical behaviour of the {Schlögl} model.
\newblock {\em Physics Letters A}, 81(4):211--214, January 1981.

\bibitem{grassberger1981montecarlo}
P.~Grassberger.
\newblock Monte {Carlo} simulations for {Schlögl}'s second model.
\newblock {\em Physics Letters A}, 84(9):459--461, August 1981.

\bibitem{grassberger1982phasetransition}
P.~Grassberger.
\newblock On phase transitions in {Schlögl}'s second model.
\newblock {\em Zeitschrift für Physik B Condensed Matter}, 47(4):365--374,
December 1982.

\bibitem{gillespie1977exact}
Daniel~T. Gillespie.
\newblock Exact stochastic simulation of coupled chemical reactions.
\newblock {\em The Journal of Physical Chemistry}, 81(25):2340--2361, December
1977.

\bibitem{shannon1948communication}
C.~E. Shannon.
\newblock A {Mathematical} {Theory} of {Communication}.
\newblock {\em Bell System Technical Journal}, 27(3):379--423, 1948.

\bibitem{tostevin2009trajectories}
Filipe Tostevin and Pieter~Rein ten Wolde.
\newblock Mutual {Information} between {Input} and {Output} {Trajectories} of
{Biochemical} {Networks}.
\newblock {\em Physical Review Letters}, 102(21):218101, May 2009.

\bibitem{meijers2019infocrit}
Matthijs Meijers, Sosuke Ito, and Pieter Rein~ten Wolde.
\newblock The behaviour of information flow near criticality.
\newblock {\em arXiv:1906.00787}, June 2019.

\bibitem{munakata2006dynamic}
T.~Munakata and M.~Kamiyabu.
\newblock Stochastic resonance in the {FitzHugh}-{Nagumo} model from a dynamic
mutual information point of view.
\newblock {\em The European Physical Journal B}, 53(2):239--243, September
2006.

 \bibitem[{\citenamefont{Alon}(2007)}]{alon2007systems}
\bibinfo{author}{\bibfnamefont{U.} \bibnamefont{Alon}},
\emph{\bibinfo{title}{An Introduction to Systems Biology: Design Principles of Biological Circuits}} (\bibinfo{publisher}{Chapman and Hall/CRC}, \bibinfo{year}{2006}).

\bibitem{skoge2011cooperativity}
Monica Skoge, Yigal Meir, and Ned~S. Wingreen.
\newblock Dynamics of {Cooperativity} in {Chemical} {Sensing} among
{Cell}-{Surface} {Receptors}.
\newblock {\em Physical Review Letters}, 107(17):178101, October 2011.

\bibitem{lopes2008spatial}
Francisco J.~P. Lopes, Fernando M.~C. Vieira, David~M. Holloway, Paulo~M.
Bisch, and Alexander~V. Spirov.
\newblock Spatial {Bistability} {Generates} hunchback {Expression} {Sharpness}
in the {Drosophila} {Embryo}.
\newblock {\em PLOS Computational Biology}, 4(9):e1000184, September 2008.

\bibitem{hudspeth2010cochlea}
A.~J. Hudspeth, Frank Jülicher, and Pascal Martin.
\newblock A {Critique} of the {Critical} {Cochlea}: {Hopf}—a
{Bifurcation}—{Is} {Better} {Than} {None}.
\newblock {\em Journal of Neurophysiology}, 104(3):1219--1229, June 2010.

\bibitem{stoop2016auditory}
Ruedi Stoop and Florian Gomez.
\newblock Auditory {Power}-{Law} {Activation} {Avalanches} {Exhibit} a
{Fundamental} {Computational} {Ground} {State}.
\newblock {\em Physical Review Letters}, 117(3):038102, July 2016.

\bibitem{friedman2006linking}
Nir Friedman, Long Cai, and X.~Sunney Xie.
\newblock Linking {Stochastic} {Dynamics} to {Population} {Distribution}: {An}
{Analytical} {Framework} of {Gene} {Expression}.
\newblock {\em Physical Review Letters}, 97(16):168302, October 2006.

\bibitem{mugler2009spectral}
Andrew Mugler, Aleksandra~M. Walczak, and Chris~H. Wiggins.
\newblock Spectral solutions to stochastic models of gene expression with
bursts and regulation.
\newblock {\em Physical Review E}, 80(4):041921, October 2009.

\bibitem{shahrezaei2008colored}
Vahid Shahrezaei, Julien~F Ollivier, and Peter~S Swain.
\newblock Colored extrinsic fluctuations and stochastic gene expression.
\newblock {\em Molecular Systems Biology}, 4(1):196, January 2008.

\bibitem{horsthemke1984noise}
W.~Horsthemke and R.~Lefever.
\newblock {\em Noise-{Induced} {Transitions}: {Theory} and {Applications} in
	{Physics}, {Chemistry}, and {Biology}}.
\newblock Springer {Series} in {Synergetics}. Springer-Verlag, Berlin
Heidelberg, 1984.

\bibitem{cotari2013cell}
Jesse~W. Cotari, Guillaume Voisinne, Orly~Even Dar, Volkan Karabacak, and
Grégoire Altan-Bonnet.
\newblock Cell-to-{Cell} {Variability} {Analysis} {Dissects} the {Plasticity}
of {Signaling} of {Common} {\textgamma} {Chain} {Cytokines} in {T} {Cells}.
\newblock {\em Science Signaling}, 6(266):ra17--ra17, March 2013.

\bibitem{bender1999asymptotics}
Carl~M. Bender and Steven~A. Orszag.
\newblock {\em Advanced {Mathematical} {Methods} for {Scientists} and
	{Engineers} {I}: {Asymptotic} {Methods} and {Perturbation} {Theory}}.
\newblock Springer-Verlag, New York, 1999.

\end{thebibliography}

\end{document}